\documentclass[%
reprint,
nofootinbib,
 amsmath,amssymb,
 aps,
]{revtex4-2}

\usepackage{physics}
\usepackage{graphicx}
\usepackage{dcolumn}
\usepackage{bm}
\usepackage{float}
\usepackage{xcolor}

\begin{document}

\preprint{APS/123-QED}

\title{{Two strategies for modeling nonlinear optics in lossy integrated photonic structures}}

\author{Milica Banic$^1$, Luca Zatti$^2$, Marco Liscidini$^2$,  J. E. Sipe$^1$}

\address{
$^1$ Department of Physics, University of Toronto, 60 St. George Street, Toronto, ON, M5S 1A7, Canada\\
$^2$ Department of Physics, University of Pavia, Via Bassi 6, 1-27100, Pavia, Italy
}

\email{mbanic@physics.utoronto.ca}

\date{\today}

\begin{abstract}
We present two complementary strategies for modeling nonlinear quantum optics in realistic {integrated optical devices}, where scattering loss is present. In the first strategy, we model scattering loss as an attenuation; in the second, we employ a Hamiltonian treatment that {includes a mechanism for scattering loss, such as a `phantom waveguide.'} {These strategies can be applied to a broad range of structures and processes. As an example, we use these two approaches to model spontaneous four-wave mixing in (i) a ring-channel system and (ii) an add-drop system.} {Even for these well-understood systems, our strategies yield some novel results. We show the rates of photon pairs, broken pairs, and lost pairs and their dependence on system parameters. We show that the properties of lost and broken photon pairs in such structures can be related to those of the un-scattered photon pairs, which are relatively simple to measure.}
\end{abstract}


\maketitle


\section{Introduction}
Integrated photonic structures are of interest both for fundamental research in nonlinear quantum optics, and for the development of platforms for quantum information processing. The generation of non-classical light in integrated structures such as waveguides and microring resonators, is now commonplace \cite{nat_comm, PhysRevLett.124.193601, PRXQuantum.2.010337_Steiner}. These structures are being employed in increasingly complex systems, due the stability of such ``on-chip" structures, and the massive integration possible for devices that consist of them \cite{Wang2020, doi:10.1126/science.aar7053, Adcock2019}. Understanding the performance of these components is {therefore essential}.

{Although one can use simple methods such as coupled-mode theory to study simple structures, such as ring resonators coupled to one or two waveguides, more sophisticated approaches are required to study nonlinear quantum optics in complex structures. For example, as we will see below, employing the asymptotic fields formalism from scattering theory is useful because it allows one to separate the linear and nonlinear dynamics of a system; once the linear behaviour is obtained, which can be done numerically for complicated structures, the nonlinear problem can easily be solved quasi-analytically. Indeed, approaches employing asymptotic field expansions have already been used in theoretical quantum optics, but these treatments have neglected the presence of loss in the structures \cite{Liscidini_asyfields, JSP, Helt:12}.}


In reality, loss due to the scattering of light off the chip is a feature that plagues integrated photonic structures. This can have both quantitative and qualitative consequences for device performance \cite{nonuniform_losses,lossy_gbs,PhysRevA.105.063707}: For example, in photon pair production, the rate at which photon pairs are generated can be reduced because of pump attenuation, and as well one of a pair of generated photons can be lost due to scattering, thereby destroying the sought-after quantum correlations between detected photons. Clearly scattering loss must be accounted for to realistically model integrated photonic structures. Here we present two complementary strategies for modeling such systems {quite generally}. 

{In Sec. \ref{section:overview} we outline the two strategies in general, and we discuss their advantages and disadvantages; in Sec. III we introduce some notation common to both. In Sec. \ref{section:firstS} we illustrate the first strategy by {applying it to a simple system; we use it to model a single ring coupled to a channel, obtaining results that are consistent with those obtained using other methods in the literature \cite{vvan,tutorial}.} In Sec. \ref{section:secondS} we do the same for the second strategy.} For example calculations we consider the generation of entangled pairs of photons by {spontaneous four-wave mixing (SFWM)}.  A deviation from the usual ``critical coupling" condition assumed to maximize the photon pair generation rate is found and discussed.  We also compare the results of the two strategies, finding good agreement for high finesse systems. In Sec. IV we use both strategies to calculate photon pair production in an add-drop structure involving a ring and two physical channels. 
{We summarize our results and present our conclusions in Sec. V.} 
{\section{The two strategies: Overview}
\label{section:overview}
}

\begin{figure}[t]
    \centering
    \includegraphics[width=0.46\textwidth]{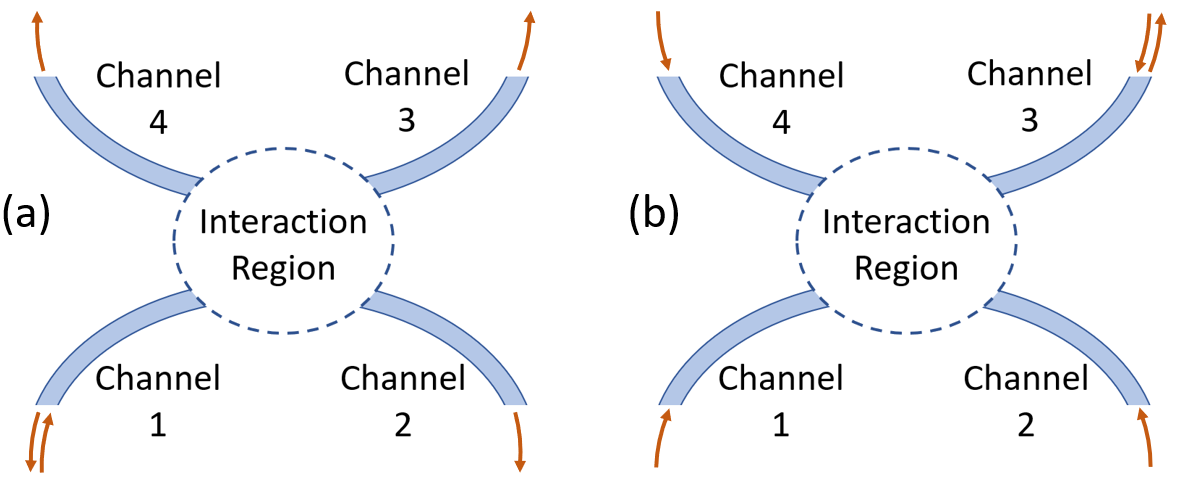}
    \caption{A sketch of {a system comprised of an interaction region with four channels. In panel (a) we represent an asymptotic-in field with respect to channel 1; in panel (b) we present an asymptotic-out field with respect to channel 3.}}
    \label{fig:InteractionRegion}
\end{figure}

{Both of the strategies {we introduce here} employ the asymptotic-in (``asy-in") and asymptotic-out (``asy-out") states familiar from scattering theory \cite{Bethe_asyfields}. {These are solutions of the linear Maxwell equations at a definite frequency and are illustrated in Fig \ref{fig:InteractionRegion}.} 
{
The asy-in fields involve propagation towards the interaction region through only one (``input") port and in general propagation out through all ports; the single incoming field has the form of a field freely propagating through the channel in the absence of any interaction region. Similarly, the asy-out fields involve propagation away from the interaction region through only one (``output") port, with the field in this region having a ``freely propagating'' form, and in general propagation towards the interaction region through all ports \cite{Liscidini_asyfields,Ge_2014,Hance2021}.} {In the absence of any truly bound states,} {either the asy-in and asy-out states can be used to expand an arbitrary field,} 
 {but the asy-in expansion is a natural choice for incoming fields (e.g. pump light), and the asy-out expansion should be used to describe fields sought at the output (e.g. light generated by nonlinear interactions).}} 

{
For realistic systems, the problem is how to calculate the asymptotic fields in the presence of scattering losses. A possible approach would be 
{to model} the system by using finite-difference-time-domain (FDTD) or finite-elements numerical tools, in which one could, in principle, take into account the structure imperfections and/or disorder that lead to scattering. Then an asy-in field would contain out-going fields not only propagating in the channels, but as well light propagating away from the chip. A similar construction holds for the asy-out fields. }

{If these full asy-in and asy-out fields were used to describe the pump and generated photons, then the resulting calculation of photon generation would include the effects of scattering. However, since even the field profile for light propagating in a scattering-free channel or ring must generally be found numerically, the full asy-in and asy-out fields would be very difficult to construct even for small systems; on top of the complexity inherent in an arbitrary lossless system, a more intricate calculation would be required to include the scattering. {This difficulty can be avoided: The two strategies we present in this paper are alternative approaches to understanding the performance of realistic, lossy integrated photonic systems.} }


{\textbf{In the first strategy}, the fields arising under linear evolution in the structure are obtained neglecting any loss; in general, this can be done numerically. Next, one includes the loss present in waveguide-like components of by introducing complex vectors associated with the propagation of the fields in those regions. As we will illustrate in Sec. \ref{section:firstS}, for asy-in fields, the imaginary part of the wavevector is simply the attenuation constant that characterizes the loss of energy as light propagates; this can be determined by a numerical calculation, or more conveniently using results of experiments in the linear regime. Correspondingly, the asy-out fields are characterized by an enhancement constant equal in magnitude to the attenuation constant of the asy-in fields.}

{Because the attenuation constant is easily determined from linear experiments on the structure, while the general form of the fields can be obtained numerically, this strategy is straightforward to implement and easily generalized {for many structures}. However, underlying this approach is the assumption that the characteristics of the scattered light are not of interest; otherwise, this phenomenological description of loss would be inadequate. While this is true in many cases, one can also envision scenarios where the properties of the scattered light need to be better understood: One might be interested in the rate of scattered photons, in the correlations between scattered and unscattered photons, or one might even consider coupling light scattered from one structure into another. For this, we turn to a different strategy.}

{\textbf{In the second strategy}, we seek to explicitly include the scattered photons in the system, but without a full numerical calculation. This can be done by modeling the scattering loss as light coupling from the lossy structure into a fictitious output channel \cite{Vernon_lossy_resonators}. With the loss modeled in this way, in principle a Hamiltonian for the system -- including the scattered photons -- can be identified, and the fields throughout the system, including regions where photons are scattered, can be identified.}

{
There are two important issues here. First, it is clearly unrealistic to apply this approach to arbitrarily complex systems. This strategy is better suited to describing simple devices, or systems composed of simple devices. Second, the approach for constructing a Hamiltonian with a mechanism for scattering depends on the structure, so this strategy is more difficult to generalize than the first strategy. However, once the second strategy is applied to a particular system, {one can investigate the properties of the scattered photons, and the full effects of scattering loss on the performance of that system {and similar ones}}.}

{We will illustrate this by applying the second strategy to a high-finesse microring resonator coupled to an arbitrary number of waveguides {(see Fig. \ref{fig:ring-channel_sketch})}. 
{For such a resonant structure,}
{scattering loss in such devices cannot be neglected due to the long dwelling time of photons in the resonator. 
{The performance of these devices is broadly relevant;} microring resonators are employed as spectral filters \cite{Barwicz:04, Ding:11}, sensors \cite{Vollmer2008,Wan:18}, and as sources of non-classical light \cite{PhysRevApplied.16.064004,PRXQuantum.2.010337_Steiner, doi:10.1126/sciadv.aba9186}, among other applications \cite{freqcomb,Joshi:16}.} 
{Moreover, this system is an example of a resonator coupled to a number of channels, and the general approach we employ can be extended to treat that large class of problems.}}

{Including a ``phantom channel" to describe loss, 
(see Fig. 5) \cite{Vernon_lossy_resonators}{, we adopt} 
a coupling model for ring-channel interactions that admits a Hamiltonian formulation and leads to an analytic solution for the fields. This coupling model relies on the assumption that the ring's finesse is high enough that the resonances are well separated; in practice {most applications aim for the high-finesse regime} \cite{Vahala2003, Yang2018, how_does_it_scale}. Indeed, since fabrication technology has advanced to the point where high-finesse resonators are readily available, this additional constraint is easily fulfilled in practice \cite{nat_comm, Yang2018, doi:10.1126/sciadv.aba9186}.}


{
We will apply the first strategy to the same type of system, both to illustrate the approach and for comparison to the second strategy. In the first strategy, we construct asy-in and -out fields only in the ring and physical waveguides. Specifically, we will use a standard approach that treats the coupling between the channel and the ring by introducing self- and cross-coupling coefficients (see Fig. \ref{fig:RingPointCoupling}{(b)}) \cite{YarivYeh}. When studying photon pair production, we can then calculate the rate at which pairs of photons will appear in the output channel, and the biphoton wave function that will characterize those pairs of photons. However, this approach does not allow us to calculate how often a pair of photons is lost to scattering or, perhaps even more importantly, how often only one photon of a generated pair is lost. These gaps will be filled by the second strategy, where the phantom channel ``keeps track'' of the lost photons.
}



{\section{Photon pair production}}
\label{section:PPR}


{Here we introduce some notation common to the two approaches.} We use the results of a quantization approach for generic integrated structures that involves the displacement field as a fundamental field operator \cite{tutorial, Quesada_QuantizeD}. It will be convenient to break up the displacement operator into its contributions from different frequency bands $J$, 
\begin{align}
    \boldsymbol{D}(\boldsymbol{r}) = \sum_J \boldsymbol{D}_J(\boldsymbol{r}), \label{eq:Dmodes}
\end{align}
where each frequency band is centered at a frequency $\omega_J$; {this will be associated with a ring resonance when we specialize to microring systems (see Fig. \ref{fig:RingPointCoupling}(a)).}

{Consider first the field in an isolated waveguide of infinite length.} Writing the $\boldsymbol{D}_J(\boldsymbol{r})$ of Eq. \eqref{eq:Dmodes} as $\boldsymbol{D}^{\text{wg}}_J(\boldsymbol{r})$, we have \cite{tutorial} 
\begin{align}
 \boldsymbol{D}^{\text{wg}}_J(\boldsymbol{r})=\int dk\boldsymbol{D}^{\text{wg}}_{Jk}(\boldsymbol{r})&a_{J}(k) +H.c.,\label{eq:D_chan_expansion}
\end{align}
where the $a_{J}(k)$
and their adjoints are ladder operators that obey the usual bosonic commutation relations, 
\begin{align}
 & \left[a_{J}(k),a_{J'}^{\dagger}(k')\right]=\delta(k-k')\delta_{JJ'},
\end{align}
provided that the light associated with each band is far from any waveguide cut-off, and is well-localized in frequency such that there is no overlap between $k$ components in distinct bands \cite{tutorial}.
We consider the $k$ of interest to range over positive values, and 
\begin{equation} \label{eq:DChannel}
\boldsymbol{D}^{\text{wg}}_{Jk}(\boldsymbol{r}) =\sqrt{\frac{\hbar \omega_{Jk}}{4 \pi}} \boldsymbol{d}_{Jk}(\boldsymbol{r_\perp)} e^{ik s} \ .
\end{equation}
Increasing $s$ 
indicates the direction in which the field is propagating, and $\boldsymbol{r_\perp}$
indicates the two Cartesian components perpendicular to that direction.  {If all the frequency bands are narrow enough that material dispersion across each one can be neglected, the $\boldsymbol{d}_{Jk}(\boldsymbol{r_\perp)}$ are normalized according to}
\begin{align}
 & \int\frac{\boldsymbol{d}_{Jk}^{*}(\boldsymbol{r_\perp})\cdot\boldsymbol{d}_{Jk}(\boldsymbol{r_\perp})}{\epsilon_{0}\varepsilon_{1}(\boldsymbol{r_\perp})}d\boldsymbol{r_\perp}=1,
\end{align}
{where $\varepsilon_{1}(\boldsymbol{r_\perp})$
is the square of the local index of refraction in frequency band $J$. This can be generalized to include material dispersion within each frequency band \cite{tutorial}.}  Finally, $\omega_{Jk}$ is the frequency of a field at $k$ in frequency range $J$; we neglect group velocity dispersion over each frequency range, and write
\begin{align}\label{eq:gv}
\omega_{Jk}=\omega_{J}+v_J(k-K_J) + ... \ ,
\end{align}
where $K_J$ is the value of $k$ at frequency $\omega_J$. Despite our neglect of material dispersion, we allow the different $v_J$ to be different, both because of modal dispersion and because the different frequency ranges identified by $\omega_J$ could also be associated with different waveguide mode profiles. We also allow $v_J$ and $K_J$ to depend on the waveguide constituting each channel, denoting the channel dependence by a superscript, $v^{(X)}_J$ and $K^{(X)}_J$. 

We now introduce the asymptotic-in and -out fields associated with a general structure such as that in Fig. \ref{fig:InteractionRegion} 
\cite{Bethe_asyfields, Liscidini_asyfields}.  
We can use Eq. \eqref{eq:Dmodes} with the $\boldsymbol{D}_J(\boldsymbol{r})$ equal to either $\boldsymbol{D}^{\text{in}}_J(\boldsymbol{r})$ or $\boldsymbol{D}^{\text{out}}_J(\boldsymbol{r})$, indicating respectively either an asymptotic-in or -out expansion, where  
\begin{align}
 \boldsymbol{D}^{\text{in(out)}}_J(\boldsymbol{r})=\sum_{X}\int dk\boldsymbol{D}_{Jk}^{\text{in(out)} (X)}(\boldsymbol{r})&a_{J}^{\text{in(out)}(X)}(k) \nonumber\\
 &+H.c.\label{eq:D_Asy_expansion}
\end{align}
The $X$ denote the different channels, and the $a_J^{\text{in(out)}(X)}(k)$ the associated ladder operators. Note that for a given channel label $X$ both $\boldsymbol{D}_{Jk}^{\text{in} (X)}(\boldsymbol{r})$ and $\boldsymbol{D}_{Jk}^{\text{out} (X)}(\boldsymbol{r})$ are in general nonvanishing for $\boldsymbol{r}$ in \emph{all} channel waveguides. 



 {Turning now to the effects of nonlinearities, we } {
consider the generation of fields through a third-order nonlinear interaction. 
described by the Hamiltonian
}
\begin{align}
    H_{\text{NL}} &= - \frac{1}{4\epsilon_0} \int d\boldsymbol{r} \Gamma_3^{ijkl}(\boldsymbol{r}) D^i(\boldsymbol{r}) D^j(\boldsymbol{r}) D^k(\boldsymbol{r}) D^l(\boldsymbol{r}) \ ,
    \label{eq:HNL1}
\end{align}
where $i,j,k,l$ are Cartesian components, summed over when repeated. The nonlinear parameter $\Gamma_3^{ijkl}(\boldsymbol{r})$ is related to the more familiar element of the third-order nonlinear tensor $\chi_{3}(\boldsymbol{r})$ by
\begin{align}
    \Gamma_{3}^{ijkl}(\boldsymbol{r}) = \frac{\chi_{3}^{ijkl}(\boldsymbol{r})}{\epsilon_0^2 \varepsilon_{1}^4(\boldsymbol{r})},
\end{align}
{where for simplicity we have assumed {the dielectric constant} $\varepsilon_1(\boldsymbol{r})$} {can be taken to be the same over all the frequency ranges of interest \cite{tutorial}.}  
If we consider only terms responsible for SFWM \cite{Sipe_Effective}, the nonlinear Hamiltonian becomes
%
%
\begin{align}
    &H_\text{SFWM} =
    \nonumber
    -\frac{1}{4\epsilon_0} \frac{4!}{2!1!1!}\\
    &\times \nonumber
    \sum_{X,X'} \int  dk_1 dk_2 dk_3 dk_4 K^{XX'}(k_1,k_2,k_3,k_4)\\ \nonumber &\times  a^{\text{out}(X)\dagger}_S(k_1) a^{\text{out}(X')\dagger}_I(k_2)\\
&\times  a_P^{\text{in}(X_{in})}(k_3) a_P^{\text{in}(X_{in})}(k_4) + H.c. \label{eq:H_SFWM1}
\end{align}

where
\begin{align}
    &K^{XX'}(k_1,k_2,k_3,k_4) =
    \nonumber \int d\boldsymbol{r} \Gamma_3^{ijkl}(\boldsymbol{r}) \\
    &\times \nonumber
    [D^{i,\text{out}(X)}_{Sk_1}(\boldsymbol{r})]^{*} [D^{j,\text{out}(X')}_{Ik_2}(\boldsymbol{r})]^{*} \\
    &\times 
    D^{k,\text{in}(X_{in})}_{Pk_3}(\boldsymbol{r})
    D^{l,\text{in}(X_{in})}_{Pk_4}(\boldsymbol{r}).
\label{eq:K}
\end{align}
We take the integral to range over the interaction region, or at least the part of it where fields can be concentrated and the nonlinear interaction is significant. Here we use $S$, $I$, and $P$ to denote respectively the signal, idler, and pump frequency ranges and modes; the combinatorial factor (4!/(2!1!1!)) takes into account that the signal and idler ranges are assumed distinct.  We have assumed that the pump fields are injected into a single channel which we label ``$X_{in}$," and in general the generated photons exit the system via different channels; we use the first superscript in $K^{XX'}(k_1,k_2,k_3,k_4)$ to denote an output channel for the signal photon, and the second an output channel for the idler photon. 

{From a standard Fermi's Golden Rule calculation with a CW pump, as detailed in Appendix \ref{appendix:new},} we find the rate of photon pair production with a signal photon in channel $X$ and an idler photon in channel $X'$ to be given by 
\begin{align}
\label{eq:Ruse}
    &R^{XX'}=\frac{72\pi^3}{\epsilon^2_0 \hbar^4 \omega^2_o} \frac{P^2_P}{v^{(X)}_S v^{(X')}_I {(v^{(X_{in})}_P)^2}} \\
    &\times \nonumber \int d\omega_1 |J^{XX'}(\omega_1,2\omega_o -\omega_1, \omega_o,\omega_o)|^2,
\end{align}
with 
\begin{align}
\label{eq:Jdef}
    J^{XX'}(\omega_1,\omega_2,&\omega_3,\omega_4) \equiv \nonumber \\ &K^{XX'}(k_1(\omega_1),k_2(\omega_2),k_3(\omega_3),k_4(\omega_4))
\end{align}
where on the right-hand side $k(\omega)$ is written using Eq. \eqref{eq:gv}.

The two strategies we introduce below differ only in the way the asymptotic-in and asymptotic-out fields that appear in $J^{XX'}(\omega_1,\omega_2,\omega_3,\omega_4)$ are constructed. {As an example, we will consider 
systems with one or more waveguides coupled to a single resonant element}; the simplest example is shown in  Fig. \ref{fig:RingPointCoupling}{(a)}, where there is a single waveguide coupled to a ring resonator {through a point coupler}. 

\section{First strategy}
\label{section:firstS}
\begin{figure}[t]
	\centering
	{\includegraphics[width=.4\textwidth]{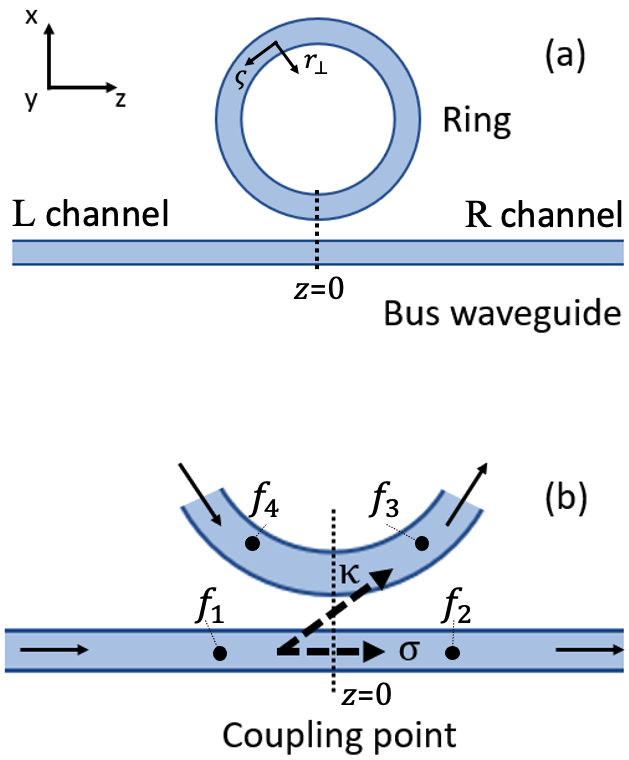}}  
	\caption{Sketch of the coupling between a bus waveguide and a ring resonator via point coupler. Coefficients $\sigma$ and $\kappa$ are the self-coupling and cross-coupling coefficients of the point coupler, respectively.}
	\label{fig:RingPointCoupling}
\end{figure}

{If we employ asymptotic-in and -out expansions for the fields when modelling a general structure, the crux of the problem is determining these fields under the linear evolution of the system; once these are determined, writing the nonlinear Hamiltonian is relatively trivial.}
In the first strategy, even very complicated structures can be treated, because the asymptotic-in and -out fields in the structure can be determined numerically.  But { to clearly illustrate the approach,}
{
here we focus on a ring-channel system as sketched in Fig. \ref{fig:RingPointCoupling}, for which analytic expressions for the fields are known \cite{YarivYeh, vvan, Helt_SFWM}. {We begin by reviewing the standard point coupling model from which the linear fields can be found. We will illustrate how the asymptotic fields are constructed from these, and how the loss is included in this approach.}}

If backscattering at the coupling point can be neglected, which is typical in well-designed systems, the ring resonator system shown in Fig. \ref{fig:RingPointCoupling}{(a)} acts as an ``all pass" filter: light incident from the left can be coupled into the ring but is eventually coupled out again into the waveguide.  Hence 
{both} 
asymptotic-in and -out fields only contain light in the waveguide propagating to the right.  {In the standard point coupling model, the two input field amplitudes ($f_1$ and $f_4$ in Fig. \ref{fig:RingPointCoupling}{(b)}) are connected to the two output field amplitudes ($f_2$ and $f_3$) by the linear system of equations}
\begin{align}\label{eq:PC_equations}
\begin{cases}
& f_2 = \sigma f_1 + i \kappa f_4 \\
& f_3 = i \kappa f_1 + \sigma f_4 \\ 
\end{cases} \ ,
\end{align}
where $\sigma$ and $\kappa$ are the self-coupling and cross-coupling coefficients of the point coupler respectively \cite{YarivYeh}; for convenience they are assumed to be real, with
\begin{align}
\label{eq:conservation}
\kappa^2+\sigma^2=1.
\end{align}
The asymptotic-in or -out field in channel $(X)$ is given by 
\begin{equation} \label{DAsyChannel}
\mathbf{D}^{\text{in(out)} (X)}_{\text{chan},Jk}(\boldsymbol{r}) =\sqrt{\frac{\hbar \omega_{Jk}}{4 \pi}} \boldsymbol{d}^{(X)}_{Jk}(x,y) f^{\text{in(out)}(X)}_{Jk}(z) e^{ik z} \ ,
\end{equation}
(cf. Eq. \eqref{eq:DChannel}), where the amplitude ${f}_{Jk}^{(X)}(z)$ takes into account the field distribution along $z$,

\begin{equation}
  {f}_{Jk}^{\text{in(out)}(X)}(z) =
    \begin{cases}
      f_1^{\text{in(out)}(X)} & \text{if $z<0$} \\
      f_2^{\text{in(out)}(X)} & \text{if $z>0$},\\
    \end{cases}       
\end{equation}
and will be different depending on whether we are specifying an asymptotic-in or asymptotic-out field. For our problem, where the pump is incident from the left, we will need asymptotic-in fields for the left $(z<0)$ channel $(X=L)$ and asymptotic-out fields for the right $(z>0)$ channel $(X=R)$. Here both channels involve the same physical waveguide, so we can put $\boldsymbol{d}^{(L)}_{Jk}(x,y)=\boldsymbol{d}^{(R)}_{Jk}(x,y) \equiv \boldsymbol{d}_{Jk}(x,y)$.

The form that these asymptotic fields take in the ring is 
\begin{equation}\label{DAsyRing}
\mathbf{D}^{\text{in(out)}}_{\text{ring},Jk}(\boldsymbol{r}) =
\sqrt{\frac{\hbar \omega_{Jk}}{4 \pi}} \boldsymbol{\mathsf{d}}_{Jk}(\boldsymbol{r}_{\perp};\zeta) f_{Jk}^{\text{in(out)}} e^{ik \zeta} \ , 
\end{equation}
where $\zeta$ is the coordinate in the direction of propagation around the ring, ranging from $0$ (at the position identified by $f_3$)
to $\mathcal{L}$ (at the position identified by $f_4$), and $\boldsymbol{r}_{\perp}$ refers to components in the plane
perpendicular to the direction indicated by increasing $\zeta$ \cite{Onodera_2016}. Thus $\boldsymbol{\mathsf{d}}_{Jk}(\boldsymbol{r}_{\perp};\zeta)$ plays the role for the ring that $\boldsymbol{d}_{Jk}(x,y)$ does for the channels.  In general it depends on all three coordinates because the direction in which the field is polarized can change with the angle $\zeta$. However, $\boldsymbol{\mathsf{d}}^*_{Jk}(\boldsymbol{r}_{\perp};\zeta) \cdot \boldsymbol{\mathsf{d}}_{Jk}(\boldsymbol{r}_{\perp};\zeta)$ will be independent of $\zeta$, and if the ring width and the width of the waveguide are taken to be the same, 
$\boldsymbol{d}_{Jk}(x,y)$ and $\boldsymbol{\mathsf{d}}_{Jk}(\boldsymbol{r}_{\perp};\zeta)$  are equal to very good approximation at $\zeta=0$.

In Fig. \ref{fig:RingPointCoupling}{(b)} $f^{\text{in(out)}}_{Jk}$ can be identified with $f_3$, and we clearly have 
\begin{align}\label{eq:Roundtrip}
f_4 = f_3 e^{ik \mathcal{L}} \ .
\end{align}
Combining Eq. \eqref{eq:PC_equations} with \eqref{eq:Roundtrip} we can then identify the asymptotic-in fields for the left channel by setting $f_1=1$, and the asymptotic-out fields for the right channel by setting $f_2=1$. We find \begin{equation}
  {f}_{Jk}^{\text{in}(L)}(z) =
    \begin{cases}
      1 & \text{if $z<0$} \\
      \frac{\sigma-e^{ik\mathcal{L}}}{1-\sigma e^{ik\mathcal{L}}} & \text{if $z>0$}\\
    \end{cases} 
    \ ,
\end{equation}
and 
\begin{align}
    {f}_{Jk}^{\text{in}}=\frac{i\kappa}{1-\sigma e^{ik\mathcal{L}}},
\end{align}
the field enhancement factor inside the ring for incident fields, while  \begin{equation}
  {f}_{Jk}^{\text{out}(R)}(z) =
    \begin{cases}
      \frac{1-\sigma e^{ik{\mathcal{L}}}}{\sigma-e^{ik\mathcal{L}}} & \text{if $z<0$} \\
      1 & \text{if $z>0$}\\
    \end{cases} 
    \ .
\end{equation}
and 
\begin{align}
    {f}_{Jk}^{\text{out}}=\frac{i\kappa}{\sigma- e^{ik\mathcal{L}}}.
\end{align}
If we use $f_{Jk}^{\text{in}}(\zeta)$ and $f_{Jk}^{\text{out}}(\zeta)$ to denote respectively the field amplitudes as they vary with $\zeta$ inside the ring (see Eq. \eqref{DAsyRing}), we have simply 
\begin{align}
\label{eq:ringfields}
    f_{Jk}^{\text{in(out)}}(\zeta)=f_{Jk}^{\text{in(out)}} e^{ik\zeta}
\end{align}
 {Note that in this simple example the asymptotic-in and -out fields differ only by a phase; for different $k$ the asymptotic-in fields $f_{Jk}^{in(L)(z)}$ have a fixed phase for $z<0$, while for different $k$ the asymptotic-out fields $f_{Jk}^{out(R)}(z)$ have a fixed phase for $z>0$ \cite{Liscidini_asyfields}.}
%
%


{Up to this point, we have done nothing to address the presence of loss in the system. For asymptotic-in fields, the} effect of scattering is a decreasing of the field intensity as light propagates; for a structure of the type considered here, that effect will be most significant inside the ring resonator. {This attenuation can be 
described phenomenologically by introducing a complex propagation wavevector}
\begin{equation} \label{ComplexWavevector}
\tilde{k}_n(\omega)=k_n(\omega) +i\frac{\xi_n(\omega)}{2}\ ,
\end{equation}
where $k_n(\omega)$ is the usual wavevector (see Eq. \eqref{eq:disp}), and $\xi_n(\omega)$ brings into effect the field intensity decay due to the propagation losses in frequency range $n$  ($\xi_3(\omega)=\xi_4(\omega)$).

For asymptotic-out states, we have $f_2=1$ instead of $f_1=1$, and in the solution of Maxwell's equations we seek there is light exiting in no other direction; the propagation in the ring is then characterized by a complex wavevector 
\begin{equation} \label{ComplexWavevector2}
\tilde{k}^{*}_n(\omega)=k_n(\omega) -i\frac{\xi_n(\omega)}{2}\ .
\end{equation}
It should be stressed that in this way one can hope to obtain the asymptotic-in and -out fields in the ring region, but without any information about the field distribution outside the structure nor about where light is scattered.
Nonetheless, following this approach and using Eq. \eqref{ComplexWavevector} and Eq. \eqref{ComplexWavevector2} in Eqs. \eqref{DAsyChannel} and \eqref{DAsyRing}, we can use Eq. \eqref{eq:Ruse} to calculate the generation rate $R^{RR}$ by applying the asymptotic fields we have identified in Eqs. \eqref{eq:K} and \eqref{eq:Jdef}. Here there is only one output channel for both the signal and the idler, so we can drop the superscripts $RR$ on $R^{RR}$, $J^{RR}$, and $K^{RR}$.  Similarly, since only one waveguide is involved, we can drop the superscripts on $v_P$, $v_S$, and $v_I$.  

\begin{figure}[t]
	\centering
	{\includegraphics[width=0.48\textwidth]{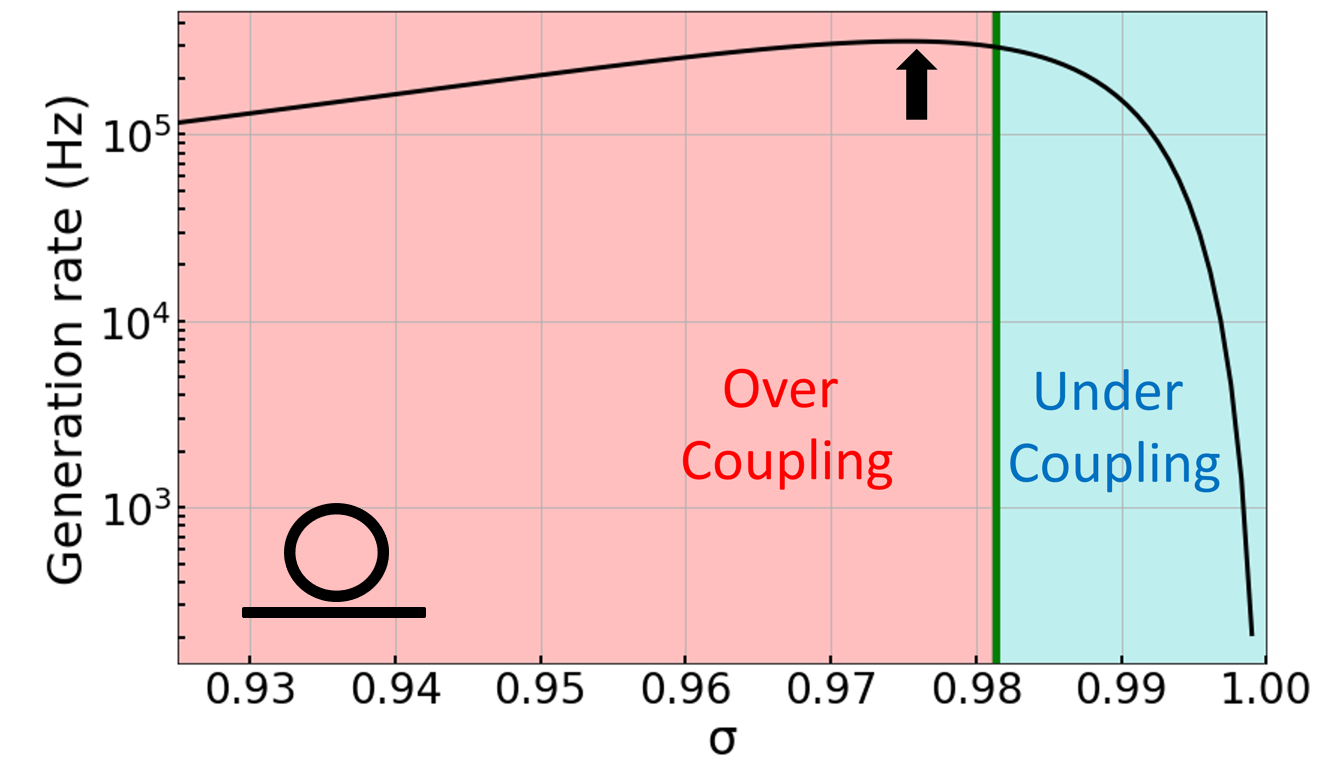}}  
	\caption{Plot of the generation rate calculated using the first approach vs the self-coupling coefficient of the point coupler. The under- and over- coupling domains are highlighted in blue and red; the green line indicates the critical coupling condition ($\sigma = a$, where $a=e^{-\xi \mathcal{L}}$ and $1-a$ is the round trip field enhancement attenuation. The black arrow indicates the maximum of the generation rate. In this simulation, the ring has a radius of $10 \, \mu \text{m}$, the loss is fixed at $\xi = 26 \, \text{dB/cm}$ $(a=0.9814, Q^{(\text{int})} \approx 2\times 10^4$). The effective index of the waveguide is $2.4$, the group velocity is $v_g = 1 \times 10^8 \, \text{m/s}$, and the nonlinear power factor is $\gamma_\text{NL}=100 \, (\text{Wm})^{-1}$.}
	\label{fig:RateRegimes}
\end{figure}

To evaluate $J(\omega_1,2\omega_o-\omega_1,\omega_o,\omega_o)$ 
we restrict the integration to the ring, where the fields will be strongest and thus the effect of the nonlinearity the greatest. A benign approximation can be made immediately, since the mode profiles $\boldsymbol{\mathsf{d}}_{Jk}(\boldsymbol{r}_{\perp};\zeta)$ are typically weak functions of $k$; we take
\begin{align}
\label{eq:dset}
    \boldsymbol{\mathsf{d}}_{Jk}(\boldsymbol{r}_{\perp};\zeta)\rightarrow \boldsymbol{\mathsf{d}}_{J}(\boldsymbol{r}_{\perp};\zeta) \equiv \boldsymbol{\mathsf{d}}_{JK_J}(\boldsymbol{r}_{\perp};\zeta).
\end{align}
 But to do a serious simplification we must assume that the $\boldsymbol{\mathsf{d}}_{J}(\boldsymbol{r}_{\perp};\zeta)$ do not depend on $\zeta$, $\boldsymbol{\mathsf{d}}_{J}(\boldsymbol{r}_{\perp};\zeta) \rightarrow \boldsymbol{\mathsf{d}}_{J}(\boldsymbol{r}_{\perp})$. This will only happen if, to good approximation, the direction of the vector field $\boldsymbol{\mathsf{d}}_{J}(\boldsymbol{r}_{\perp};\zeta)$ points everywhere ``normal to the chip."  Only in this limit do the integrations over $\boldsymbol{r}_{\perp}$ and $\zeta$ in $J(\omega_1,2 \omega_o-\omega_1,\omega_o,\omega_o)$ factor to two separate integrals. Introducing a coefficient characterizing the nonlinearity,   
\begin{align}
\label{eq:gammaNL}
    &\gamma_{\text{NL}} =
    \\ &\frac{3 \omega_P}{4\epsilon_0 v_P^2} \int d\boldsymbol{r}_{\perp} \Gamma_3^{ijkl}(\boldsymbol{r}_{\perp})   \mathsf{d}^{*i}_{S}(\boldsymbol{r}_{\perp}) \mathsf{d}^{*j}_{I}(\boldsymbol{r}_{\perp}) \mathsf{d}^{k}_{P}(\boldsymbol{r}_{\perp}) \mathsf{d}^{l}_{P}(\boldsymbol{r}_{\perp}), \nonumber
\end{align}
which involves an integral only over $\boldsymbol{r}_{\perp}$, from Eq. \eqref{eq:Ruse} we find a generation rate of photon pairs given by  
\begin{align}\label{RateNum}
R =& \frac{ 1}{2 \pi} \left(\frac{|\gamma_\text{NL}| P_P}{\omega_P} \right)^2 \frac{v^2_P}{v_S v_I} \int d\omega_1 \omega_1 (2\omega_o - \omega_1) \nonumber \\
& \times \left|  \mathcal{J}(\omega_1, 2\omega_o-\omega_1, \omega_o, \omega_o) \right|^2\ .
\end{align}
Here $\mathcal{J}(\omega_1,2\omega_o-\omega_1,\omega_o,\omega_o)$ captures the effect of the variation over the ring of the fields in Eq. \eqref{eq:ringfields}; namely,
\begin{align}
    &\mathcal{J}(\omega_1,\omega_2,\omega_3,\omega_4) =
    \\ & \int(f^{\text{out}}_{S\tilde{k}_1^*(\omega_1)}(\zeta) f^{\text{out}}_{I\tilde{k}_2^*(\omega_2)}(\zeta))^*f^{\text{in}}_{P\tilde{k}_3(\omega_3)}(\zeta)f^{\text{in}}_{P\tilde{k}_4(\omega_4)}(\zeta) d \zeta
    \nonumber
    \\& =(f^{\text{out}}_{S\tilde{k}_1^*(\omega_1)} f^{\text{out}}_{I\tilde{k}_2^*(\omega_2)})^*f^{\text{in}}_{P\tilde{k}_3(\omega_3)}f^{\text{in}}_{P\tilde{k}_4(\omega_4)} \frac{e^{i(\Delta \tilde{k}) \mathcal{L}}-1}{i(\Delta \tilde{k})},
    \nonumber
\end{align}
since the integral over $\zeta$ runs from $0$ to $\mathcal{L}$, with 
\begin{align}
    \Delta \tilde{k} = \tilde{k}_3(\omega_3)+\tilde{k}_4(\omega_4)-\tilde{k}_1(\omega_1)-\tilde{k}_2(\omega_2).
\end{align}
In Fig. \ref{fig:RateRegimes} we plot the generation rate of photon pairs as a function of the coupling constant $\sigma$, assuming the $\xi_j(\omega)$ are the same for all $j$ and independent of frequency, and assuming that indeed the vector field $\boldsymbol{\mathsf{d}}_{J}(\boldsymbol{r}_{\perp};\zeta)$ points everywhere normal to the chip so that the reductions made above are valid. We find that although the maximum of the intensity inside a ring resonator is known to be reached at critical coupling, the maximum of the generation rate occurs when the system is slightly over coupled.

%
%
%
\section{Second Strategy} 
\label{section:secondS}

%
%

{In the second strategy, we define a linear Hamiltonian for the structure which explicitly includes a loss mechanism; the asymptotic fields for the lossy structure are determined by referring to the equations of motion produced by this Hamiltonian.} {We illustrate this for a ring resonator coupled to an arbitrary number of waveguides, one of these being a ``phantom waveguide", which we introduce to model scattering losses from the resonator \cite{Vernon_lossy_resonators}; with this, the asymptotic fields will explicitly include the scattered photons.} {This approach is more complex than others that have been used to model this type of system, but unlike other approaches, this strategy makes it simple to study the properties of the scattered photons, and to consider an arbitrary number of physical bus waveguides.}


\subsection{Fields and Hamiltonian}

\begin{figure}[t]
    \centering
    \includegraphics[width=.48\textwidth]{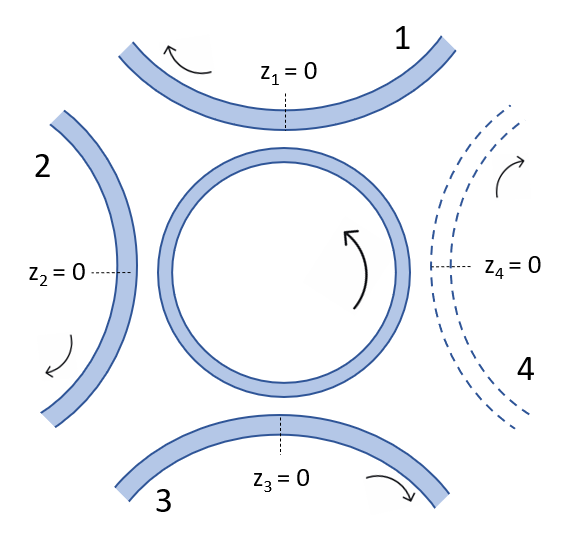}
    \caption{A sketch of a ring-waveguide system for N=4 waveguides. Solid lines indicate physical waveguides; the dotted line indicates the phantom waveguide.}
    \label{fig:ring-channel_sketch}
\end{figure}

{We begin by outlining the free propagation of fields in the waveguides and in the ring, and the method used to describe the coupling between the ring and a given waveguide.} For each waveguide field, as introduced in Sec. \ref{section:PPR}, we take the coordinate $z$ to label the direction of propagation, with the components of $\boldsymbol{r}_{\perp}$ then being $x$ and $y$. As indicated in Fig \ref{fig:ring-channel_sketch}, for each waveguide the coupling point with the ring is located at $z=0$ for that particular waveguide. We introduce the terms `input region' and `output region' to refer to the parts of the waveguide before and after the coupling point, respectively. The `input region' is then defined by $z<0$, and the `output region' by $z>0$. We introduce 
\begin{align}
 & \psi_{J}(z)=\int\frac{dk}{\sqrt{2\pi}}a_{J}(k)e^{i(k-K_{J})z},
\end{align}
and 
the linear Hamiltonian for an isolated waveguide can be written as 
\begin{align}
 H_{L}^{\text{wg}}&=\sum_{J}\hbar\omega_{J}\int\psi_{J}^{\dagger}(z)\psi_{J}(z)dz\label{eq:HLchan}\\ \nonumber &-\frac{1}{2}i\hbar v_{J}\int\left(\psi_{J}^{\dagger}(z)\frac{\partial\psi_{J}(z)}{\partial z}-\frac{\partial\psi_{J}^{\dagger}(z)}{\partial z}\psi_{J}(z)\right)dz,
\end{align}
provided the frequency bands are narrow enough that to good approximation we can put $\hbar \omega_{Jk} \approx \hbar \omega_J$, and $\boldsymbol{d}_{Jk}(x,y) \approx \boldsymbol{d}_{J}(x,y) \equiv \boldsymbol{d}_{JK_J}(x,y)$ \cite{tutorial}. Under this approximation, we also have from Eq. \eqref{eq:D_chan_expansion}
\begin{align}
     & \boldsymbol{D}_J(\boldsymbol{r})=\sqrt{\frac{\hbar\omega_{J}}{2}} \boldsymbol{d}_{J}(x,y) e^{iK_J z} \psi_{J}(z).
     \label{eq:DJ}
\end{align}

Using our earlier definition of $\boldsymbol{\mathsf{d}}_{J}(\boldsymbol{r}_{\perp};\zeta)$ in Eq.  \eqref{eq:dset}, the field in an isolated ring can be written as \cite{tutorial}
\begin{align}
 & \boldsymbol{D}(\boldsymbol{r})=\sum_{J}\sqrt{\frac{\hbar\omega_{J}}{2}}\frac{\boldsymbol{\mathsf{d}}_{J}(\boldsymbol{r}_{\perp};\zeta)}{\sqrt{\mathcal{L}}} b_J e^{i\kappa_{J}\zeta}+H.c.,
\end{align}
where
\begin{align}
 & \left[b_{J},b_{J'}^{\dagger}\right]=\delta_{JJ'},
\end{align}
with
$\kappa_{J}=2\pi m_{J}/\mathcal{L}$,
where $m_{J}$ is the index of the mode. The linear Hamiltonian for the ring modes is 
\begin{align}
 & H_{L}^{\text{ring}}=\sum_{J}\hbar\omega_{J}b_{J}^{\dagger}b_{J}.\label{eq:HLring}
\end{align}
Finally, we treat the coupling between the ring and a waveguide by introducing different coupling coefficients associated with each mode of the isolated ring. {Unlike in the standard point-coupling model described in Sec. \ref{section:firstS}, here we model the coupling with}
\begin{align}
 & H_{L}^{\text{coupling}}=\sum_{J}\left(\hbar\gamma_{J}b_{J}^{\dagger}\psi_{J}(0)+H.c.\right)\label{eq:HLcoupling},
\end{align}
{where $\gamma_{J}$ is a constant characterizing the strength of the coupling between the discrete ring mode and the continuous channel mode at frequency band $J$.} {Eq. \eqref{eq:HLcoupling} is valid provided the ring resonances are well separated (high finesse), so that these distinct modes $J$ are well-defined and can be identified \cite{Vernon_lossy_resonators}.} This is an important distinction between the two strategies; the standard point-coupling model used in Sec. \ref{section:firstS} is not constrained by this assumption, and can describe low-finesse structures. In practice, the high-finesse regime is usually of interest, {and Eq. \eqref{eq:HLcoupling} is appropriate \cite{Vahala2003, Yang2018, how_does_it_scale}.}

{One can introduce}
\begin{align}
    \Gamma_J = \frac{|\gamma_J|^2}{2v_J} \label{eq:Gamma_def},
\end{align}
{where we will see that $\Gamma_J$ characterizes the rate at which intensity in the ring can decay as it couples into the waveguide; if the waveguide were coupled only to a single waveguide, $\Gamma_J$ would identify the resonance linewidth. In the high-finesse limit, this parameter can be related to the self-coupling constant $\sigma$ in the standard point-coupling model used in Sec. \ref{section:firstS} by} 
\begin{align}
    \Gamma_J = \frac{(1-\sigma)v_J}{\mathcal{L}}
\end{align}
\cite{tutorial}, under the assumption that the ring and the waveguide are made of the same material, have the same cross-section, and thus in particular the same $v_J$; but this can be generalized.


%
%

As in the simpler scenario of Fig. \ref{fig:RingPointCoupling}{(a)}, in the scenario of Fig \ref{fig:ring-channel_sketch} there are twice as many channels as waveguides. For the direction of light propagation indicated, half of the channels will have asymptotic-in states of interest associated with them (those in the region of their waveguide with $z<0$), and we call them ``in-channels." The other half will have asymptotic-out states of interest associated with them (those in the region of their waveguide with $z>0$), and we call them ``out-channels." We denote the waveguide with which channel $X$ is associated by $\bar X$.  

With an arbitrary number of waveguides, {one being a phantom waveguide}, the linear Hamiltonian is given by 
\begin{align} \label{eq:HL}
 & H_{L}= H_{L}^{\text{ring}}+\sum^\prime_X H_{L}^{\text{wg}{(\bar X)}}+\sum^\prime_X H_{L}^{\text{coupling}{(\bar X)}} , 
\end{align}
where the prime on the sums indicates a sum over in-channels, resulting in a sum over waveguides, and the terms in the Hamiltonian are given by 
(\ref{eq:HLchan}), (\ref{eq:HLring}), and (\ref{eq:HLcoupling}) respectively for each waveguide; for the waveguide $\bar X$ associated with an in-channel $X$ we introduce a set of operators $\psi_J^{(X)}(z)$, group velocities $v_J^{(X)}$, and coupling constants $\gamma_J^{(X)}$. {The associated decay rate due to each channel $X$ is labelled $\Gamma_J^{(X)}$ (related to $\gamma_J^{(X)}$ by Eq. \eqref{eq:Gamma_def}), and resonator's total linewidth is given by}
\begin{align}
    \bar{\Gamma}_J = \sum_X \Gamma_J^{(X)} \label{eq:Gammabar_def}.
\end{align}
{The total linewidth, and each waveguide's contribution to it, can be expressed in terms of quality factors as}
\begin{align}
    \bar{\Gamma}_J &= \frac{\omega_J}{2 Q^{(\text{load})}_J} \label{eq:Qload}\\
    \Gamma^{(X)}_J &= \frac{\omega_J}{2 Q^{(X)}_J}\label{Qcoupling},
\end{align}
{where $Q^{(\text{load})}_J$ is the loaded quality factor, and $Q^{(X)}_J$ is the `coupling' quality factor associated with each waveguide; for example, if $X$ labels the phantom waveguide, $Q^{(X)}_J$ is the resonator's intrinsic quality factor. We will also refer to the escape efficiency from the ring into channel X, defined as}
\begin{align}
    \eta_J^{(X)} = \frac{\Gamma^{(X)}_J}{\bar{\Gamma}_J}. \label{eq:escape_efficiency}
\end{align}

\subsection{Asymptotic fields}



To derive the asymptotic field amplitudes, we use the Heisenberg equation with (\ref{eq:HL}) to find the equations of motion that the fields must satisfy, and we impose the appropriate boundary conditions for the asymptotic fields; the details of the calculation are given in Appendix \ref{appendix:asy_amplitudes}. We find that for an in-channel $X$ the associated asymptotic-in field amplitudes are given by 

\begin{align}
 \boldsymbol{D}&_{Jk}^{\text{in}(X)}(\boldsymbol{r})=\sqrt{\frac{\hbar\omega_{J}}{4\pi}}\boldsymbol{d}^{(X)}_{J}(x,y)e^{ikz},\label{eq:in_amplitudes} \\ \nonumber &\qquad\boldsymbol{r}\in\text{input region of waveguide $\bar X$,}\\ \nonumber
 &=0,\\ \nonumber &\qquad\boldsymbol{r}\in\text{input region of all other waveguides}\\ \nonumber
 &=-\sqrt{\frac{\hbar\omega_{J}}{4\pi}}\boldsymbol{\mathsf{d}}_{J}(\boldsymbol{r}_{\perp};\zeta)e^{i\kappa_{J}\zeta}F_{J-}^{(X)}(k),\\ \nonumber
 &\qquad\boldsymbol{r}\in\text{ring},\\ \nonumber
 &=\sqrt{\frac{\hbar\omega_{J}}{4\pi}}\boldsymbol{d}^{(X)}_{J}(x,y)\left(1+\sqrt{\mathcal{L}}\frac{i\gamma^{(X)}_{J}}{v^{(X)}_{J}}F^{(X)}_{J-}(k)\right)e^{ikz},\\ \nonumber &\qquad\boldsymbol{r}\in\text{output region of waveguide $\bar X$,}\\ \nonumber
 &= \sqrt{\frac{\hbar\omega_{J}}{4\pi}}\boldsymbol{d}^{(Y)}_{J}(x,y)\left(\frac{i\gamma^{(Y)}_{J}}{v^{(Y)}_{J}}\sqrt{\mathcal{L}}F^{(X)}_{J-}(k)\right)e^{ikz} ,\\ \nonumber
 &\qquad\boldsymbol{r}\in\text{output region of all other waveguides $\bar Y \neq \bar X$.}
\end{align}
We have introduced 
\begin{align}
    F^{(X)}_{J\pm}(k) = \frac{1}{\sqrt{\mathcal{L}}} \left(\frac{\left(\gamma_{J}^{(X)}\right)^{*}}{v^{(X)}_J\left(K^{(X)}_J - k \right) \pm i \bar{\Gamma}_J}\right), \label{eq:F}
\end{align}
the complex field enhancement factors that arise in this resonant structure, linking the field in the ring to the input in asymptotic-in fields (see Eq. \eqref{eq:in_amplitudes} above), and the field in the ring to the output in asymptotic-out fields (see Eq. \eqref{eq:out-amplitude} below).

For an out-channel $X$ the associated asymptotic-out field amplitudes are given by
\begin{align}
 \boldsymbol{D}&_{Jk}^{\text{out}(X)}(\boldsymbol{r})=\sqrt{\frac{\hbar\omega_{J}}{4\pi}}\boldsymbol{d}^{(X)}_{J}(x,y)e^{ikz}, \label{eq:out-amplitude}\\ \nonumber &\qquad\boldsymbol{r}\in\text{output region of waveguide $\bar X$,}\\ \nonumber
 &=0,\\ \nonumber &\qquad\boldsymbol{r}\in\text{output region of all other waveguides,}\\ \nonumber
 &=-\sqrt{\frac{\hbar\omega_{J}}{4\pi}}\boldsymbol{\mathsf{d}}_{J}(\boldsymbol{r}_{\perp};\zeta)e^{i\kappa_{J}\zeta}F^{(X)}_{J+}(k),\\ \nonumber&\qquad\boldsymbol{r}\in\text{ring},\\ \nonumber
 &=\sqrt{\frac{\hbar\omega_{J}}{4\pi}}\boldsymbol{d}_{J}^{(X)}(x,y)\left(1-\sqrt{\mathcal{L}}\frac{i\gamma_{J}^{(X)}}{v^{(X)}_{J}}F^{(X)}_{J+}(k)\right)e^{ikz},\\ \nonumber&\qquad\boldsymbol{r}\in\text{input region of waveguide $\bar X$,}\\ \nonumber
 &=\sqrt{\frac{\hbar\omega_{J}}{4\pi}}\boldsymbol{d}_{J}^{(Y)}(x,y)\left(-\sqrt{\mathcal{L}}\frac{i\gamma^{(Y)}_{J}}{v^{(Y)}_{J}}F^{(X)}_{J+}(k)\right)e^{ikz},\\ \nonumber&\qquad\boldsymbol{r}\in\text{input region of all other waveguides $\bar Y \neq \bar X$.}
\end{align}

\subsection{Spontaneous four-wave mixing}
\label{section:SFWM_analytic}

{With the asymptotic-in and -out fields in hand, restricting as usual the integral in Eq. \eqref{eq:K} to range over the ring we find}
\begin{align}\label{eq:KXX'ring}
    K^{XX'}&(k_1,k_2,k_3,k_4) = \frac{\hbar^2 \epsilon_0 \left(v^{(X_{in})}_P\right)^2}{12\pi^2} \sqrt{\omega_S \omega_I} \,\overline{\gamma}_{NL}\, \mathcal{L} \nonumber \\ \times & F^{(X)*}_{S+}(k_1) F^{(X')*}_{I+}(k_2) F^{(in)}_{P-}(k_3) F^{(in)}_{P-}(k_4),
\end{align}
where we have 
introduced the nonlinear parameter
\begin{align}
    &\overline\gamma_{\text{NL}} = \frac{3 \omega_P}{4\epsilon_0 \left(v_P^{(in)}\right)^2 \mathcal{L}} \int d\boldsymbol{r}_{\perp}d\zeta \Gamma^{ijkl}(\boldsymbol{r}_{\perp}) \\&\times  \mathsf{d}^{*i}_{S}(\boldsymbol{r}_{\perp};\zeta) \mathsf{d}^{*j}_{I}(\boldsymbol{r}_{\perp};\zeta) \mathsf{d}^{k}_{P}(\boldsymbol{r}_{\perp};\zeta) \mathsf{d}^{l}_{P}(\boldsymbol{r}_{\perp};\zeta)e^{i\Delta \kappa \zeta}, \nonumber
\end{align}
and we have introduced $\Delta \kappa = 2\kappa_P - \kappa_S - \kappa_I$, the wavenumber mismatch in the ring.  Note that if the polarization of the profiles $\boldsymbol{\mathsf{d}}_{P}(\boldsymbol{r}_{\perp};\zeta)$ are everywhere ``normal to the chip," and if $(\Delta \kappa) \mathcal{L} \ll 1$ -- the latter can be expected if the signal, idler, and pump resonances are closely spaced -- then we have $\overline \gamma_{\text{NL}} \rightarrow \gamma_{\text{NL}}$ (see Eq. \eqref{eq:gammaNL}). 
Putting Eq. {\eqref{eq:KXX'ring}} in \eqref{eq:Ruse}, we find
\begin{align}
    {R}^{XX'} &=  \label{eq:asyfields-rate} \frac{\sqrt{\omega_S \omega_I}}{\omega_o} \frac{\left(v_P^{(X_{in})}\right)^2}{v^{(X)}_S v^{(X')}_I} (\bar \gamma_{NL} \mathcal{L})^2  \frac{1}{\hbar \omega_o} P_P^2 P_{\text{vac}}\\&\times  |F^{(in)}_{P-}(k_o)|^4 |F^{(X)}_{S+}(K_S)|^2 |F^{(X')}_{I+}(K_I)|^2, \nonumber 
\end{align}
where again $k_o$ and $\omega_o$ are the wavenumber and frequency of the CW pump, which can be detuned from resonance. We have identified
\begin{align}
    P_{\text{vac}} &= \frac{\hbar}{2}  \frac{ \sqrt{\omega_S \omega_I}\, \overline{\Gamma}_S\overline{\Gamma}_I(\overline{\Gamma}_S+\overline{\Gamma}_I)}{(2\omega_o - \omega_S - \omega_I)^2 + (\overline{\Gamma}_S+\overline{\Gamma}_I)^2},
\end{align}
the fluctuating vacuum power \cite{how_does_it_scale}. {This can be written in terms of the ring's loaded quality factors using \eqref{eq:Qload}; we will do this for the sample calculation in Sec. \ref{section:ring-channel}}. 
{The}  quantity is independent of the channels through which the photons exit; {this is expected, since here the vacuum power only relates to the generation of the photons in the resonator.} 

By inspecting Eq. \eqref{eq:asyfields-rate} and recalling the definition of $F^{(X)}_{J\pm}(k)$ in Eq. \eqref{eq:F}, we see that the rate of generation of signal and idler photons in any pair of channels $X$ and $X'$ can be related to the rate of generation of signal and idler photons in another pair of channels $Y$ and $Y'$ by
\begin{align}
    \frac{{R}^{XX'}}{{R}^{YY'}} = \frac{\Gamma_S^{(X)}}{\Gamma_S^{(Y)}} \frac{\Gamma_I^{(X')}}{\Gamma_I^{(Y')}} = \frac{\eta_S^{(X)} \eta_I^{(X')}}{\eta_S^{(Y)} \eta_I^{(Y')}}, \label{eq:rate_relative}
\end{align}
{where the escape efficiency $\eta_J^{(X)}$ is defined in \eqref{eq:escape_efficiency}. With \eqref{eq:rate_relative}, one can infer the rates in all sets of channels, given the rate of photons in one arbitrary pair of channels, and the relative coupling constants between the channels and the ring. For example, one could take $Y=Y'$ as the pair of channels where the rate is known, so that the rate could be approximated by coincidence measurements in channel $Y$. From this quantity the rates of broken and scattered pairs can be inferred.} 

If we consider an incident pump pulse sufficiently weak that at most a pair of photons are generated, we can calculate the biphoton wave function (the joint spectral amplitude (JSA)) that characterizes the pair; we again use the interaction Hamiltonian defined in Eq. \eqref{eq:H_SFWM1}, but in taking the limit of a classical pump we use $a^{(X_{in})}_P(k) \rightarrow \alpha \phi(k)$, where $\alpha$ is the field amplitude and the pulse function $\phi(k)$ is normalized according to 
\begin{align}
    \int |\phi(k)|^2 dk = 1.
\end{align}
Eq. \eqref{eq:H_SFWM1}, in the interaction picture, then becomes 
\begin{align}
&H^{(I)}_{\text{SFWM}}(t) =\nonumber - \frac{\hbar^2}{4\pi^2}  \left(v^{(X_{in})}_P\right)^2 \sqrt{\omega_S \omega_I} \bar\gamma_{\text{NL}} \mathcal{L} \alpha^2 \sum_{X,X'} \nonumber \\ \nonumber&\times \int dk_1 dk_2 dk_3 dk_4 e^{-i(\omega_{P}(k_4) + \omega_{P}(k_3) - \omega_{I}(k_2) - \omega_{S}(k_1))t} \nonumber \\ &\times F^{(X)*}_{S+}(k_1) F^{(X')*}_{I+}(k_2) F^{(X_{in})}_{P-}(k_3) F^{(X_{in})}_{P-}(k_4) \phi(k_3)  \nonumber \\ &\times \phi(k_4)  a^{(X)\dagger}_S(k_1) a^{(X')\dagger}_I(k_2)+ H.c.
\end{align}
To first order, the ket that results after the pump pulse passes is  
\begin{align}
    \ket{\Psi} &\approx \ket{\text{vac}} - \frac{i}{\hbar} \int^{\infty}_{-\infty} dt' H_{SFWM}^{(I)}(t')\ket{\text{vac}}+...
    \label{eq:ket}\\
    &= \ket{\text{vac}} + \beta \sum_{X,X'} \ket{XX'} + ...  \label{eq:pair_ket}
\end{align}
(see Appendix \ref{appendix:new}), with
\begin{align}
    \nonumber \ket{XX'} = &\int dk_1 dk_2 \phi^{XX'}(k_1,k_2)\\&\times a_S^{(X)\dagger}(k_1) a_I^{(X')\dagger}(k_2) \ket{\text{vac}},
\end{align}
where again the first superscript $X$ referring to the signal output channel and the second superscript $X'$ to the idler output channel; $\phi^{XX'}(k_1,k_2)$ is a biphoton wave function, normalized according to 
\begin{align}
    \sum_{XX'} \int dk_1 dk_2 |\phi^{XX'}(k_1,k_2)|^2 = 1. \label{eq:BWF_normalization}
\end{align}
We assume that $\beta \ll 1$ such that higher order terms are negligible, and so $|\beta|^2$ is approximately the probability that a pair of photons is generated. We have 
%
%
%
\begin{align}
    \phi^{XX'}(k_1,k_2) &= \nonumber \frac{\alpha^2}{\beta} \frac{i\hbar}{2\pi} \sqrt{\omega_S \omega_I} \left(v_P^{(X_{in})}\right)^2 \bar\gamma_{\text{NL}} \mathcal{L}\\&\times F^{(X)*}_{S+}(k_1) F^{(X')*}_{I+}(k_2)g(k_1,k_2), \label{eq:biphoton_2}
\end{align}
where 
\begin{align}
    \nonumber g(k_1,k_2) &= \int dk_3 dk_4  F^{(X_{in})}_{P-}(k_3) F^{(X_{in})}_{P-}(k_4) \phi(k_3) \phi(k_4) \\&\times
    \delta(\omega_{P}(k_4) + \omega_{P}(k_3) - \omega_{I}(k_2) - \omega_{S}(k_1))
\end{align}
is determined by the shape of the pump pulse, {and $|\beta|^2$ can be determined by using Eq. \eqref{eq:BWF_normalization} with \eqref{eq:biphoton_2}}.

Using Eq. \eqref{eq:biphoton_2}, we can relate the joint spectral amplitude associated with a pair of channels $X,X'$ to the JSA associated with another pair of channels $Y,Y'$; recognizing that the field enhancement factor in Eq. \eqref{eq:F} can be rewritten using
\begin{align}
    \omega(k) = \omega_J +  v^{(X)}_J\left(k - K^{(X)}_J \right),
\end{align}
we can see that 
\begin{align}
    \phi^{XX'}(k_1,k_2) &= \frac{\gamma_S^{(X)} \gamma_I^{(X')}}{\gamma_S^{(Y)} \gamma_I^{(Y')}} \phi^{YY'}(k_1,k_2), \label{eq:biphoton_relative}
\end{align}
{which can be written in terms of more familiar parameters using Eqs. \eqref{eq:Gamma_def}, \eqref{eq:Qload}, and  \eqref{eq:escape_efficiency}.}

As we discussed for Eq. \eqref{eq:rate_relative_2}, here again one can define a `reference' pair of channels for which the biphoton wave function is known, and infer all the other terms of the biphoton wave function using {escape efficiencies and group velocities for the different channels}. {For example, the biphoton wave function associated with lost pairs and broken pairs can be inferred from the JSA associated with photons pairs in a particular channel; the latter can be obtained either from a simpler model such as the first strategy, or from coincidence measurements.}

From Eq. \eqref{eq:biphoton_relative} we see that the shape of the biphoton wave function associated with each pair of channels is the same; this in turn implies that for this type of system, scattering has no effect on the system's full biphoton wavefunction beyond its contribution to the total linewidth $\bar{\Gamma}_J$. This is easily understood: Since the nonlinear effects are confined to the ring, and the ring-channel coupling is frequency independent, the spectral properties of the photon pairs do not depend on which channels they couple into. The amplitude associated with each pair of channels depends on the coupling constants $\gamma_J^{(X)}$, as expected.

\subsection{Ring-channel system}
\label{section:ring-channel}

\begin{figure}[h]
    \centering
    \includegraphics[width=.48\textwidth]{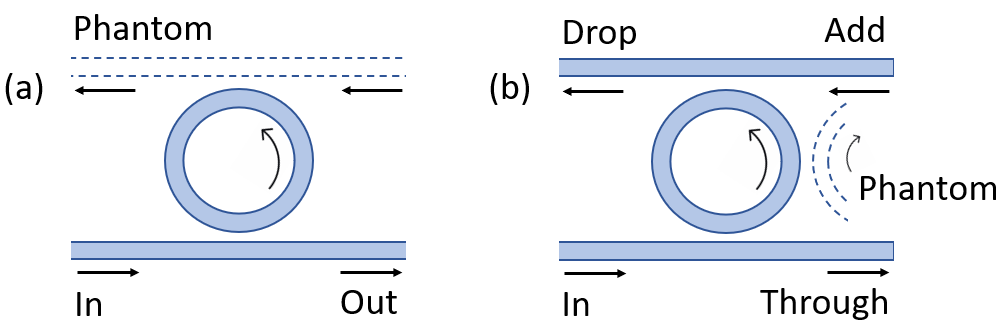}
    \caption{Sketches of (a) a ring-channel system and (b) an add-drop system. In both cases, a phantom channel is included to account for scattering losses when using the second approach.}
    \label{fig:ring+phantom}
\end{figure}

We now apply these general results to the simple case of a lossy ring coupled to a bus waveguide (Fig. \ref{fig:ring+phantom}{(a)}). Since we model the ring's scattering loss as a coupling to a phantom waveguide, here the labels $X$ and $X'$ range over two channels \cite{Vernon_lossy_resonators}. 

We use the label $O$ to denote the out-channel of the bus waveguide, and $\mathcal{P}$ to denote the out-channel of the phantom waveguide, as indicated in Fig. \ref{fig:ring+phantom}{(a)}; photons that exit via channel $O$ can be detected, whereas photons that exit via channel $\mathcal{P}$ are scattered and lost. We then have four scenarios, each with an associated rate: Both photons can appear at the output ($R^{OO}$), the signal can appear at the output while the idler is lost ($R^{O\mathcal{P}}$), the idler can appear at the output while the signal is lost ($R^{\mathcal{P} O}$), or both photons can be lost ($R^{\mathcal{P} \mathcal{P}}$). We predict the rate of pairs arriving at the output to be 
\begin{align}
    {R}^{OO} &= \label{eq:R_SOIO} \frac{\sqrt{\omega_S \omega_I}}{\omega_P} \frac{\left(v_P^{(X_{in})}\right)^2}{v^{(X)}_S v^{(X')}_I} (\bar \gamma_{\text{NL}} \mathcal{L})^2  \frac{1}{\hbar \omega_P} P_P^2 P_{\text{vac}} \\&\times |F^{(X{in})}_{P-}(K_P + \delta K_P)|^4 |F^{(O)}_{S+}(K_S)|^2 |F^{(O)}_{I+}(K_I)|^2,\nonumber
\end{align}
Using Eq. \eqref{eq:rate_relative} we have
\begin{align}
    \label{eq:rate_relative_2} {R}^{O\mathcal{P}} &=  \left(\frac{1-\eta_I}{\eta_I}\right){R}^{OO}, \\ \nonumber
    {R}^{\mathcal{P} O} &=  \left(\frac{1-\eta_S}{\eta_S}\right) {R}^{OO}, \\ \nonumber
    {R}^{\mathcal{P} \mathcal{P}} &= \left(\frac{1-\eta_S}{\eta_S}\right) \left(\frac{1-\eta_I}{\eta_I}\right) {R}^{OO}.
\end{align}
{In Eq. \eqref{eq:rate_relative_2} we refer only to the escape efficiency for the physical channel $\eta_J \equiv \eta^{(O)}_J$; there are only two channels, so $\eta^{(P)}_J = 1-\eta^{(O)}_J$.} All the rates are equal at critical coupling, where $\eta_J = 0.5$. 

We now provide a sample calculation for the rates, assuming system parameters that are compatible with current fabrication technology, and consistent with those used in Sec. \ref{section:firstS}. We assume the pump to be on resonance, and we assume that $2\omega_P \approx (\omega_S + \omega_I)$. Then we have 
\begin{align}
    {R}^{OO} &= \nonumber \frac{\sqrt{\omega_S \omega_I}}{\omega_P} \frac{\left(v_P^{(X_{in})}\right)^2}{v^{(X)}_S v^{(X')}_I}  (\bar \gamma_{\text{NL}} \mathcal{L})^2  \frac{1}{\hbar \omega_P} P_P^2 P_{\text{vac}}  \\&\times|F^{(X_{in})}_{P-}(K_P)|^4 |F^{(O)}_{S+}(K_S)|^2 |F^{(O)}_{I+}(K_I)|^2, \label{eq:R_SOIO2}
\end{align}
with 
\begin{align}\label{eq:Pvac_Q}
    P_{\text{vac}} &= \frac{\hbar}{2}  \sqrt{\omega_S \omega_I}   \frac{\overline{\Gamma}_S\overline{\Gamma}_I}{ (\overline{\Gamma}_S+\overline{\Gamma}_I)}\\ \nonumber
    &= \frac{\hbar}{4} \frac{{(\omega_S \omega_I)}^{3/2}}{ \omega_S Q^{(\text{load})}_I + \omega_I Q^{(\text{load})}_S},
\end{align}
{
For example, with $Q^{(\text{int})}_I \approx Q^{(\text{int})}_S = 2\times10^4$, $\eta_I \approx \eta_S = 0.5$, and $\lambda_I \approx \lambda_S = 1550$ nm, we have $Q^{(\text{load})}_I \approx Q^{(\text{load})}_S = 1\times10^4$ and $P_{\text{vac}} = 1.9$ nW.} 


For the field enhancement factors, having taken the pump to be on resonance we have
\begin{align}
    |F_{J\pm}^{(X)}(K_J)|^2 &= \frac{1}{\mathcal{L}} \frac{|\gamma^{(X)}_J|^2}{\bar{\Gamma}_J^2}\\ \nonumber
    &= \frac{4 v_J^{(X)} \eta_J Q^{(\text{load})}_J}{2\pi \mathcal{R} \omega_J},
\end{align}
where $\mathcal{R}$ is the radius of the ring. Taking $v_J^{(X)} = 10^8$ m/s 
and $\mathcal{R} = 10 \mu$m, we have $|F_{J\pm}^{(X)}(K_J)|^2 = 26.2$. We assume that the resonances in question are spectrally close, such that these parameters are representative of each resonance, so that $|F_{P\pm}^{(X)}(K_P)|^2 \approx |F_{S\pm}^{(X)}(K_S)|^2 \approx |F_{I\pm}^{(X)}(K_I)|^2 \approx$ 26.2. Taking $\bar\gamma_{\text{NL}} =\gamma_{\text{NL}}  = 100$ (Wm)$^{-1}$ and $P_P = 1$ mW, we estimate $R^{\mathcal{OO}} = 1.08\times10^4$ pairs per second. {And from Eq. \eqref{eq:rate_relative_2}, {at critical coupling} we expect the same rate for each of the three other possible trajectories
.}

\begin{figure}[H]
	\centering
	{\includegraphics[width=.48\textwidth]{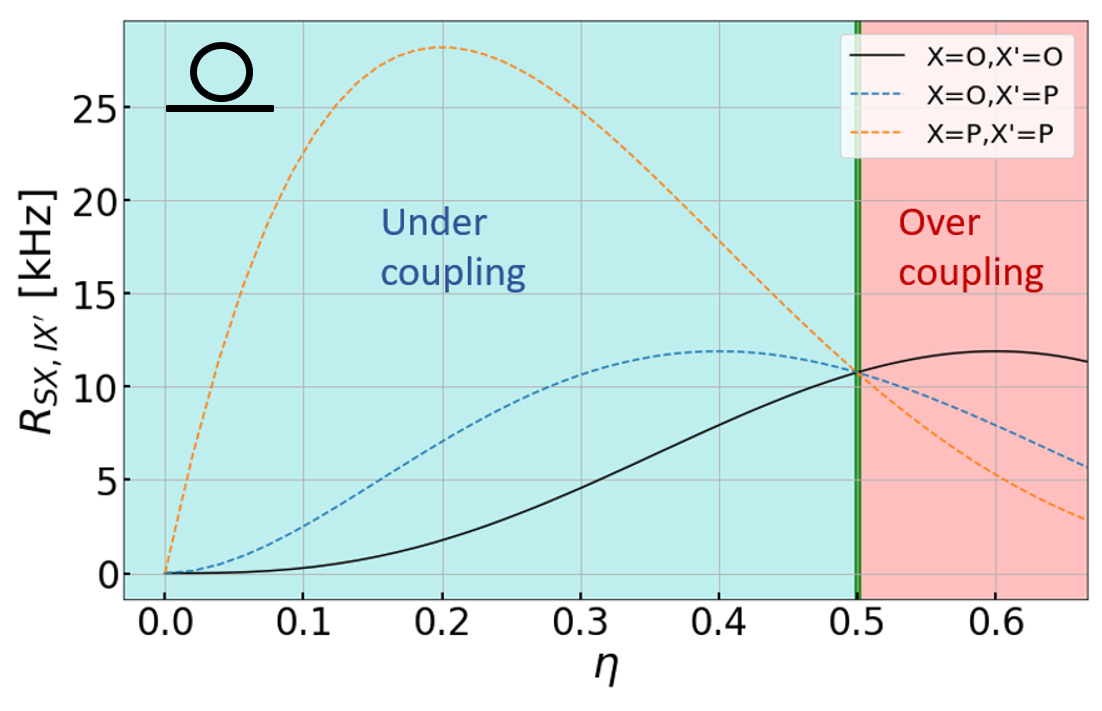}}  
	\caption{Rates of photon pairs exiting via each pair of channels as a function of the escape efficiency $\eta$. Here the $X=P,X'=O$ configuration is omitted because it is identical to the $X=O,X'=P$ case.}
	\label{fig:ring-channel_rates}
\end{figure}

\begin{figure}[h]
	\centering
	{\includegraphics[width=.48\textwidth]{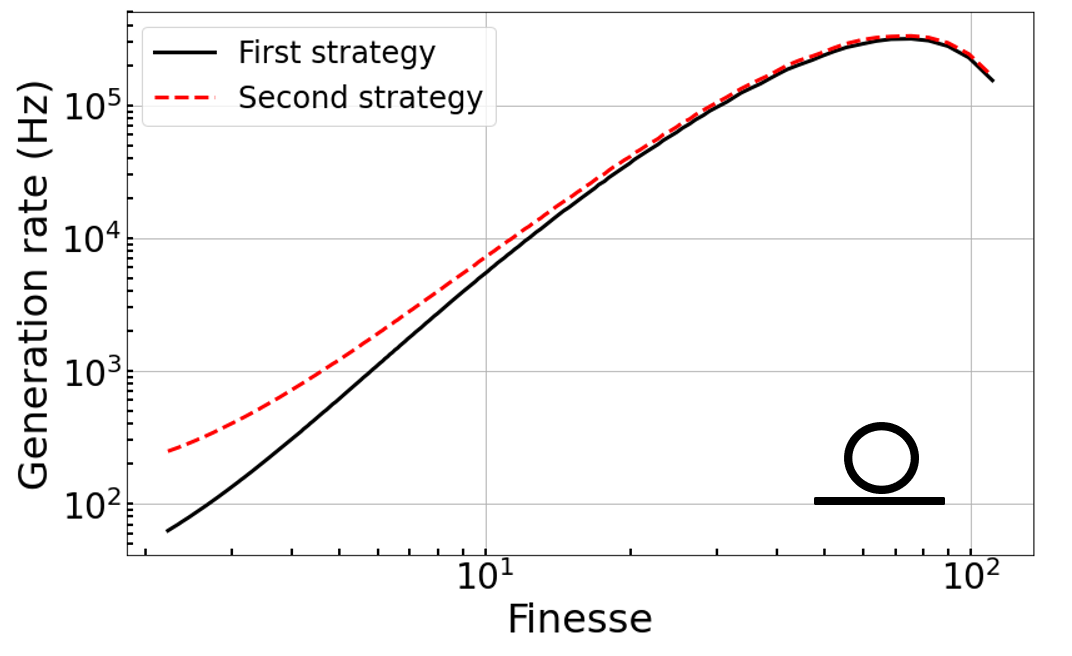}} 
	\caption{Pair generation rate calculated using the first strategy (black solid line) and the second strategy (red dashed line) for a ring coupled to a single channel, as a function of resonator finesse. The parameters summarized in Fig. \ref{fig:RateRegimes} are again used here.}
	\label{fig:RateFinesseRing}
\end{figure}

In Fig. \ref{fig:ring-channel_rates}, we plot the rate of photon pairs exiting from each pair of channels as the ring-channel coupling is increased, assuming the system parameters listed above. The rates are equal at the critical coupling point. As noted in Sec. \ref{section:firstS}, the rate of photon pairs at the output is maximized with the system slightly overcoupled ($\eta = 0.6$); here we see that the other rates decrease as the coupling is increased past the critical coupling point. A slightly overcoupled system is thus favourable in two ways: first, the rate of pairs at the output is maximized, and second, the ratio of unbroken pairs to broken pairs is larger. In Fig. \ref{fig:RateFinesseRing} we compare these analytic results to numerical results obtained as described in Sec. \ref{section:firstS}. As expected, the two methods agree well when the ring's finesse is sufficiently large. 



\begin{figure}[h]
    \centering
    \includegraphics[width=0.48\textwidth]{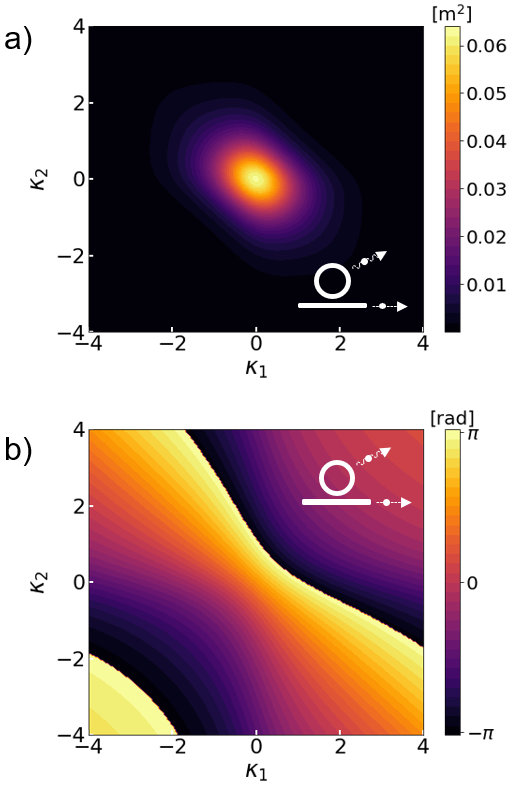}
    \caption{{Plots of (a) the modulus squared and (b) the phase of one term of the biphoton JSA; the plot here represents the signal photon coupling out, with the idler being scattered  (see Eq. \eqref{eq:biphoton_2}). We have used the normalization in Eq. \eqref{eq:BWF_normalization}, and the axes are in terms of dimensionless variables $\kappa_{1(2)} = v_{S(I)}(k_{1(2)}-K_{S(I)})/\overline{\Gamma}_{S(I)}$.} We assume a 10 ps Gaussian pulse, and we have set $\eta = 0.6$. The ring parameters are consistent with those used in Fig. \ref{fig:RateRegimes}.}
    \label{fig:ring-channel_JSA}
\end{figure}

Finally, in Fig. \ref{fig:ring-channel_JSA} we plot the JSA for the ring-channel system. We note that the JSA components associated with different pairs of channels differ only by a constant amplitude and a global phase, as discussed in Sec. \ref{section:SFWM_analytic}. Even when the terms associated with broken and lost pairs are included in the full biphoton wavefunction, its shape has the form familiar from earlier treatments and experiments which address only the photon pairs at the output \cite{JSP,Helt_SFWM}.

\section{Sample calculation: add-drop}
\label{section:add-drop}

We now use the two strategies to model photon pair generation via SFWM in an add-drop system, sketched in Fig. \ref{fig:ring+phantom}{(b)}. We take the input fields to be entering by the channel labeled `in', although the channel labeled `add' could also be used as an input channel. The generated photons can couple from the ring into the {`through'} channel ($\mathcal{T}$), the {`drop'} channel ($\mathcal{D}$), or they can be scattered; in the second strategy, this is modeled as a coupling into the phantom channel ($\mathcal{P}$). There are thus nine trajectories that the generated pairs can take; the first strategy can describe the four trajectories that involve neither photon being scattered (solid lines in Fig. \ref{fig:add-drop_rates}), while the second strategy can describe all of them. 

\begin{figure}[H]
	\centering
	{\includegraphics[width=.48\textwidth]{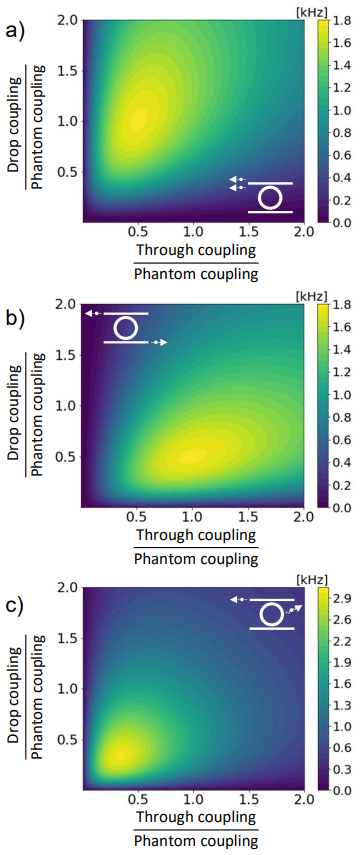}}  
	\caption{The dependence of pair rates on the through and drop channel coupling ($\Gamma^{(\mathcal{T})}$ and $\Gamma^{(\mathcal{D})}$, respectively). Of the nine configurations, here we plot the subset in which the signal photon exits via the drop channel; the idler photon couples into the drop, through, and phantom channel in panels (a), (b), and (c) respectively.}
	\label{fig:add-drop_contours}
\end{figure}

\begin{figure}[h]
	\centering
	{\includegraphics[width=.48\textwidth]{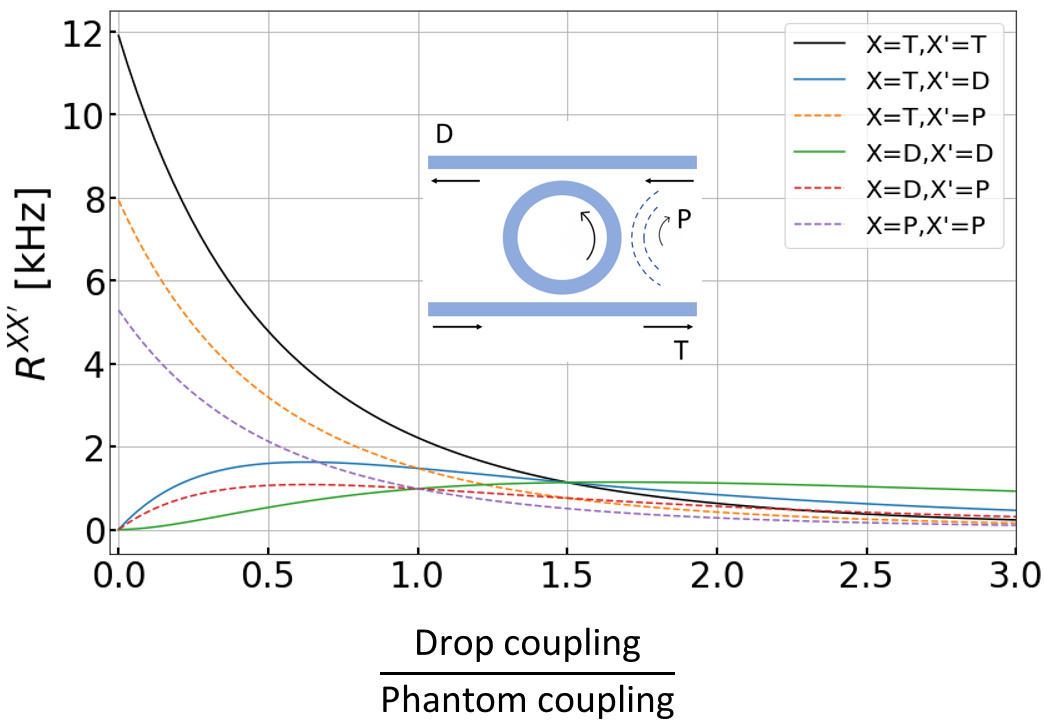}}  
	\caption{Rates associated with each photon pair trajectory with increasing drop channel coupling. Since the signal and idler frequencies are similar, we have $\Gamma^{(X)}_S \approx \Gamma^{(X)}_I$ and $R^{XX'} = R^{X'X}$, so three of the nine rates are omitted from the legend and can be inferred from the others.}
	\label{fig:add-drop_rates}
\end{figure}

{The rate of photons exiting in each set of channels depends on the resonator's coupling to the through and drop channels.} The effect of varying these two parameters on the distribution of the generated pairs is not obvious; on top of the trade-off between coupling and field enhancement (seen in Fig. \ref{fig:add-drop_rates} and discussed below), we have variable coupling to two channels which affect the rates in different ways, since only the through channel carries the pump field. In Fig. \ref{fig:add-drop_contours} we plot the rates associated with a subset of the configurations; the full set of plots can be found in Appendix \ref{section:supplementary}. 

{For example, consider a scenario where one wishes to have the photon pairs coupling into the drop channel; Fig. \ref{fig:add-drop_contours}{(a)} indicates that to maximize the rate of photon pairs in the drop channel, the drop coupling should be set equal to the phantom coupling (which is set by $Q^{(\text{int})}$), with the through coupling smaller by roughly a factor of two.} 

One can also envision a more nuanced case: One might wish to have photon pairs coupling into the drop channel, and to suppress the rates at which only one photon couples into the drop channel, while the other photon couples into the through channel (see Fig. \ref{fig:add-drop_contours}{(b)}) or is scattered (Fig. \ref{fig:add-drop_contours}{(c)}). Using Fig. \ref{fig:add-drop_contours} one can see the trade-off: One could reduce the rate of photons in the unwanted trajectories (b and c) by increasing the drop coupling and/or decreasing the through coupling with respect to the values that maximize the rate of pairs in the drop channel.

For easier interpretation we can consider a slice of the parameter space in Fig. \ref{fig:add-drop_contours} where the coupling to one of the waveguides is constant. In Fig. \ref{fig:add-drop_rates} we plot the rates of all the trajectories in the case where the through coupling is fixed such that {$\eta^{(\mathcal{T})}_J = 1.5 \eta^{(\mathcal{P})}_J$, where $\eta^{(\mathcal{P})}_J$ is set by $Q^{(\text{int})}$; here again we use $Q^{(\text{int})} = 2\times10^4$}.

\begin{figure}[h]
	\centering
	{\includegraphics[width=.48\textwidth]{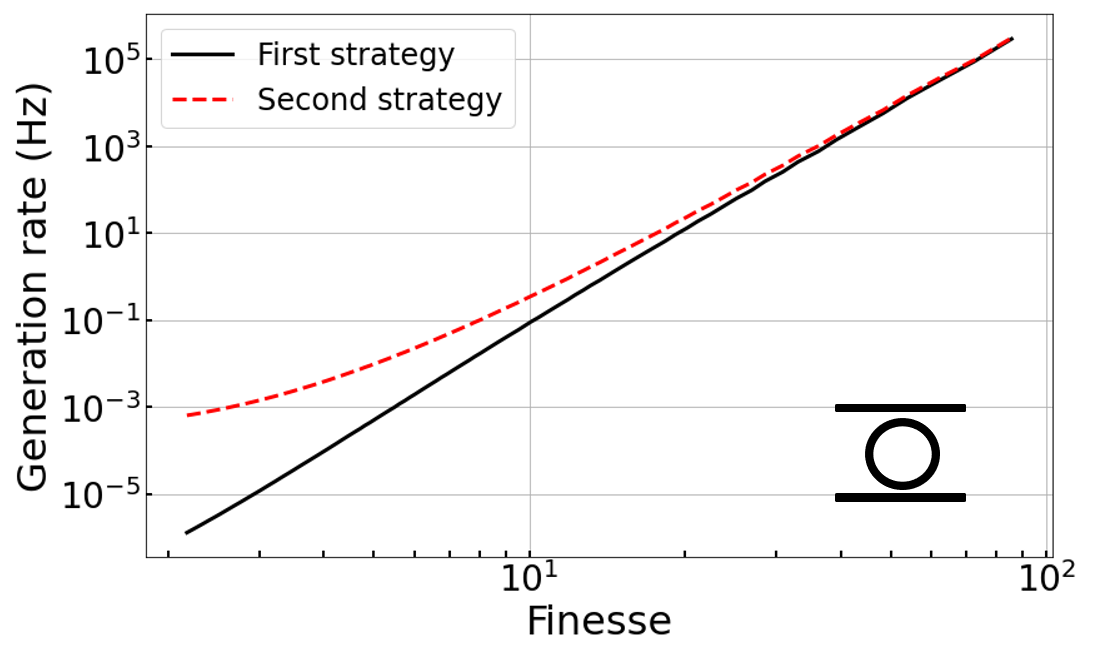}}  
	\caption{Pair generation rate using the first strategy (black solid line) and the second strategy (red dashed line) for an add-drop system, as a function of resonator finesse. The parameters summarized in Fig. \ref{fig:RateRegimes} are again used here. We fix the coupling between the ring and the in/through channel ($\sigma_1=0.9814$) and vary the coupling with the add/drop channel ($\sigma_2 \in [0.3,1]$). Here the pair generation rate refers to the number of photon pairs per unit time collected in the through channel.}
	\label{fig:RateFinesseAddDrop}
\end{figure}

The behaviour shown in Fig. \ref{fig:add-drop_rates} aligns well with intuition: the increased coupling to the drop channel is initially accompanied by increasing rates of photons coupling into the drop channel. As the coupling is increased past a critical point
, the rates decrease; increasing the coupling to the drop channel increases the total linewidth of the resonator. This results in a lower field enhancement, and the total pair generation rate falls. We again compare the rates predicted by the first and second strategies in Fig. \ref{fig:RateFinesseAddDrop}, and find that the second strategy is valid provided the ring's finesse is sufficiently large. 


It is straightforward to plot the JSA for the add-drop system, but we omit it since it is qualitatively identical to the ring-channel JSA given in Fig. \ref{fig:ring-channel_JSA}. The two systems differ only in the coupling constants 
, which affects the scaling, and in their full linewidths 
which affects the shape of the JSA in a trivial way (recall Eqs. \eqref{eq:biphoton_2} and \eqref{eq:biphoton_relative}).  

\section{Conclusion}
We have developed two strategies for realistic modeling of integrated photonic systems under the influence of scattering loss. In the first strategy, the fields  {must be determined -- usually numerically -- directly from the system's structure, and} 
the attenuation 
{due to} scattering loss is modeled with a complex wavevector. It is easy to implement and generalize, but it describes only photons which are not scattered, providing only a partial understanding of the effect of scattering loss. In the second strategy, the fields are obtained starting from a Hamiltonian for the system, which includes a loss mechanism, such as a fictitious channel, into which the scattered photons couple. Because the scattered photons are accounted for, this strategy provides a more complete understanding of the effect of scattering loss, but it is {more} difficult to generalize to arbitrary structures. 



{By applying the second strategy to a microring system, we find that the characteristics of scattered photons can be inferred from the characteristics of photon pairs which couple out of the system (see Eqs. \eqref{eq:rate_relative} and \eqref{eq:biphoton_relative}). The latter behaviour can be understood either using a simple model like the first strategy, or from experimental observations. We also show that the presence of scattering loss in this type of system has a trivial effect on the correlations of unscattered photons. These results should be immediately useful in the interpretation of experimental results from realistic microring systems, or from larger systems that include them, and in the design of new systems. We emphasize that this conclusion applies only to the simple microring system we have considered, or other similar resonant systems; further efforts could focus on generalizing the second approach to different types of systems, in which less trivial behaviour could arise. }

{We have also limited our analysis to the generation of photon pairs by a classical pump, but applying this approach to other scenarios could yield other interesting and useful results. For example, it would be straightforward to apply the strategies discussed here to study the generation of squeezed light. With the second strategy one could also consider different types of light as the input to a microring system; an arbitrary state of light can be considered the input. We envision this approach being applied to a number of other problems. One could study the effect of linear dynamics on exotic states of light, or study nonlinear optics driven by non-classical light, such as squeezed light or a very weak coherent state such that non-Gaussian interactions could be important.}

\begin{acknowledgments}
 M.B. acknowledges support from the University of Toronto Faculty of Arts \& Science Top Doctoral Fellowship. J.E.S. and M.B. acknowledge support from the Natural Sciences and Engineering Research Council of Canada.  M.L. and L.Z. acknowledge support by Ministero dell’Istruzione, dell’ Università e della Ricerca (Dipartimenti di Eccellenza Program (2018–2022)).
\end{acknowledgments}

\newpage


\bibliography{apssamp}

\onecolumngrid

\clearpage
\appendix

\section {Technical details}
\label{appendix:new}

{Here we give the details of the calculation leading to Eq. \eqref{eq:Ruse} for the pair production rate from SFWM in the CW limit. We begin by moving into an interaction picture with Eq. \eqref{eq:H_SFWM1} as the perturbation. The linear Hamiltonian describing free propagation of the fields leads to the raising and lowering operators in Eq. \eqref{eq:H_SFWM1} all acquiring the usual time dependences. Then treating the strong pump field classically, we can put $a_P^{\text{in}(X_{in})}(k)\rightarrow \alpha_P(k)$, where $\alpha_P(k)$ is a complex function. We assume a CW pump at a frequency $\omega_o$, which can be slightly detuned from the center frequency of the pump range. Then $\omega_o=\omega_P+\delta\omega_P$, with the associated $k$ given to first order by $k_o=K^{(X_{in})}_P+(\omega_o-\omega_P)/v^{(X_{in})}_P$\text (recall Eq. \eqref{eq:gv}), and we have
\begin{align}
\label{eq:alphaPset}
    \alpha_P(k)=\sqrt{\frac{2 \pi P_P}{\hbar\omega_o v^{(X_{in})}_P}}\delta(k-k_o).
\end{align}
where $P_P$ is the pump power.}

This follows from noting first that in the asymptotic-in fields the form of the incident field for channel $X_{in}$ is the same as it would be for an infinite channel, and so we can identify $\alpha_P(k)$ by considering an infinite channel where the energy would be 
\begin{align}
    &H=\int dk \hbar \omega_{Pk} \alpha^*_P(k) \alpha_P(k),
\end{align}
where we take the integral over all $k$, although for our applications $\alpha_P(k)$ will be nonzero only for positive $k$. Moving to position space and taking 
\begin{align}
\label{eq:phiAppdef}
    &\phi_P(z)=\int \frac{dk}{\sqrt{2 \pi}} \alpha_P(k) e^{i(k-k_o)z},
\end{align}
for a wave packet with components $k$ only very near $k_o$ we have
\begin{align}
    &H \approx \hbar \omega_o \int dk \alpha^*_P(k) \alpha_P(k) \\
    \nonumber
    & = \hbar \omega_o \int dz \phi^*_P(z) \phi_P(z)
\end{align}
Thus the local energy density is $\hbar\omega_o\phi^*_P(z) \phi_P(z)$, and since the group velocity is $v_P$ the local power is $\hbar \omega_o v^{(X_{in})}_P \phi^*_P(z) \phi_P(z)$. To set this equal to $P_P$ we see from Eq. \eqref{eq:phiAppdef} that we should indeed set $\alpha_P(k)$ according to Eq. \eqref{eq:alphaPset}. 

{Putting \eqref{eq:alphaPset} into the interaction picture version of Eq. \eqref{eq:H_SFWM1}, we have 
\begin{align}
   H^I_{\text{SFWM}}(t) = &
   -\sum_{X,X'} \int dk_1 dk_2 M^{XX'}(k_1,k_2)e^{-i\Omega(k_1,k_2) t}
   a^{\text{out}(X)\dagger}_S(k_1) a^{\text{out}(X')\dagger}_I(k_2),
\end{align}
where 
\begin{align}
    &M^{XX'}(k_1,k_2)=\frac{6\pi P_P}{\epsilon_0\hbar\omega_o v_P}K^{XX'}(k_1,k_2,k_o,k_o), 
    \label{eq:MXX'}
\end{align}
and
\begin{align}
\label{eq:Omega}
    &\Omega(k_1,k_2)=2\omega_o - \omega_{Sk_1}-\omega_{Ik_2}.
\end{align}
}

To estimate the pair production rate we begin with Schr\"{o}dinger's equation in the interaction picture,
\begin{align}
 & i\hbar\frac{d}{dt}\left|\Psi\left(t\right)\right\rangle =H_\text{SFWM}^{(I)}(t)\left|\Psi(t)\right\rangle ,\label{eq:Schroedinger}
\end{align}
where as usual $H_\text{SFWM}^{I}(t) = e^{i H_0 t/\hbar}H_\text{SFWM}e^{-i H_0 t/\hbar}$ and $H_0$ is the full linear Hamiltonian. If we assume that the classical pump field is on from $-T/2$ to $T/2$, and that the initial state has no signal or idler photons, the iterative solution of Eq. \eqref{eq:Schroedinger} is 

\begin{align}
 & \left|\Psi\left(t>\frac{T}{2}\right)\right\rangle =\left|\text{vac}\right\rangle - \frac{i}{\hbar}\int_{-\frac{T}{2}}^{\frac{T}{2}}H_\text{SFWM}^{(I)}(t)\left|\text{vac}\right\rangle dt+...\label{eq:psi_evolve}
\end{align}
The second term in this iteration is a two-photon state, and so to first order, the probability of generating a photon pair is 

\begin{align}
 & \mathcal{P}=\frac{1}{\hbar^{2}}\int_{-\frac{T}{2}}^{\frac{T}{2}}dt\int_{-\frac{T}{2}}^{\frac{T}{2}}dt'\bra{\text{vac}} H_\text{SFWM}^{(I)}(t')H_\text{SFWM}^{(I)}(t)\ket{\text{vac}}, \label{eq:Pwork}
\end{align}
and from Eq. \eqref{eq:MXX'} we have 

\begin{align}
    \bra{\text{vac}} H_\text{SFWM}^{(I)}(t')H_\text{SFWM}^{(I)}(t)\ket{\text{vac}} = \sum_{X,X'} \int dk_1 dk_2 |M^{XX'}(k_1,k_2)|^2 e^{-i\Omega(k_1,k_2)(t-t')}, 
\end{align}
where we have used $[a^{(X)}_{J}(k),a^{\dagger(Y)}_{J}(k')] = \delta_{XY}\delta(k-k')$, and $\Omega(k_1,k_2)$ is given by Eq. \eqref{eq:Omega}. Putting this into Eq. \eqref{eq:Pwork} and evaluating the integrals over time, we have 

\begin{align}
 & \mathcal{P}=\frac{1}{\hbar^{2}}\sum_{X,X'}\int dk_1 dk_2 \frac{4 \text{sin}^2 \left(\Omega(k_1,k_2) T/2\right)}{\Omega^2(k_1,k_2)} |M^{XX'}(k_1,k_2)|^2.
\end{align}
We then assume that $T$ is sufficiently long that we can use

\begin{align}
 \frac{4 \text{sin}^2 \left(\Omega(k_1,k_2) T/2\right)}{\Omega^2(k_1,k_2)} \rightarrow 2\pi T \delta(\Omega(k_1,k_2)),
\end{align}
so that 

\begin{align}
 & \mathcal{P}=\frac{2\pi T}{\hbar^{2}}\sum_{X,X'}\int dk_1 dk_2 \delta(\Omega(k_1,k_2)) |M^{XX'}(k_1,k_2)|^2.
\end{align}
The total pair generation rate is then 
\begin{align}
    R = \frac{\mathcal{P}}{T} = \frac{2\pi}{\hbar^{2}}\sum_{X,X'}\int dk_1 dk_2 \delta(\Omega(k_1,k_2)) |M^{XX'}(k_1,k_2)|^2 = \sum_{X,X'} R^{XX'},
\end{align}
where the rate of pair production with a signal photon in channel $X$ and an idler photon in channel $X'$ is given by 

\begin{align}
    \label{eq:RXX'}
    R^{XX'} = \frac{2\pi}{\hbar^{2}}\int dk_1 dk_2 \delta(\Omega(k_1,k_2)) |M^{XX'}(k_1,k_2)|^2,
\end{align}
which is Eq. \eqref{eq:RXX'}.

{Moving from wave number $(k_1,k_2)$ to frequency variables $(\omega_1 \equiv \omega_{Sk_1}, \omega_2 \equiv \omega_{Ik_2})$, and putting 
\begin{align}
    J^{XX'}(\omega_1,\omega_2,&\omega_3,\omega_4) \equiv K^{XX'}(k_1(\omega_1),k_2(\omega_2),k_3(\omega_3),k_4(\omega_4))
\end{align}
where on the right-hand side 
\begin{align}
\label{eq:disp}
&k_1(\omega)=K^{(X)}_S+\frac{\omega-\omega_S}{v^{(X)}_S}, \\
\nonumber
&k_2(\omega)=K^{(X')}_I+\frac{\omega-\omega_I}{v^{(X')}_I}, \\
\nonumber
&k_3(\omega)=k_4(\omega)=K^{(X_{in})}_P+\frac{\omega-\omega_P}{v^{(X_{in})}_P}, 
\end{align}
using  Eq. \eqref{eq:MXX'} in \eqref{eq:RXX'}, and implementing the Dirac delta function, we find
\begin{align}
    &R^{XX'}=\frac{72\pi^3}{\epsilon^2_0 \hbar^4 \omega^2_o} \frac{P^2_P}{v^{(X)}_S v^{(X')}_I {(v^{(X_{in})}_P)^2}} \int d\omega_1 |J^{XX'}(\omega_1,2\omega_o -\omega_1, \omega_o,\omega_o)|^2,
\end{align}
which is Eq. \eqref{eq:Ruse}.}

\section{Deriving asymptotic field amplitudes for a ring coupled to many waveguides}
\label{appendix:asy_amplitudes}

{First, recall that as indicated in Fig \ref{fig:ring-channel_sketch}, there are twice as many channels as waveguides. Each waveguide has an ``in-channel" and an ``out-channel", in the region of the waveguide with $z<0$ and $z<0$ respectively, for the direction of propagation indicated in Fig \ref{fig:ring-channel_sketch}. The in-channels have asymptotic-in states of interest associated with them, while the out-channels have asymptotic-out states of interest associated with them. We associate the label $X$ with a particular channel, and we denote the corresponding waveguide by $\bar X$.}  

{As described in Sec. \ref{section:overview}, an asymptotic-in field for a particular channel $X$ generally consists of a field propagating towards an interaction region in only that channel, and outgoing fields in all other channels. The incoming field has the form that it would have if it were freely propagating in that channel. This applies to a ring-channel system as follows: At negative enough $z$ in the appropriate waveguide (outside of the interaction region with the ring), an asymptotic-in field for a ring-channel system will have the form of the field that would propagate were the ring not present \cite{Liscidini_asyfields}. Then for an asymptotic-in field associated with
an in-channel $X$ we can identify 
\begin{align}
 & \boldsymbol{D}_{Jk}^{\text{in}(X)}(\boldsymbol{r})=\boldsymbol{D}_{Jk}^{\text{wg}(\bar X)}(\boldsymbol{r})\label{eq:link} \\
 & \text{(in input region of waveguide $\bar X$),} \nonumber \\
 & \boldsymbol{D}_{Jk}^{\text{in}(X)}(\boldsymbol{r})= 0 \label{eq:linka}\\
 & \text{(in input region of other waveguides $\bar Y$),} 
 \nonumber 
\end{align}
where $\boldsymbol{D}_{Jk}^{\text{in}(X)}(\boldsymbol{r})$ is the asymptotic-in field amplitude, as introduced in  Eq. \eqref{eq:D_Asy_expansion}. Similarly, an asymptotic-out field consists of a single outgoing field in one channel, and fields incoming from every channel. At positive enough $z$ in the appropriate waveguide, an asymptotic-out field in a ring-channel system will have the form of the field that would propagate were the ring not present; for the asymptotic-out field associated with an out-channel $X$, we can identify 
\begin{align}
 & \boldsymbol{D}_{Jk}^{\text{out}(X)}(\boldsymbol{r})=\boldsymbol{D}_{Jk}^{\text{wg}(\bar X)}(\boldsymbol{r})\label{eq:link2} \\
 & \text{(in output region of waveguide $\bar X$),} \nonumber \\
 & \boldsymbol{D}_{Jk}^{\text{out}(X)}(\boldsymbol{r})= 0 \label{eq:link2a}\\
 & \text{(in output region of other waveguides $\bar Y$).} 
 \nonumber 
\end{align}
}

{To find the asymptotic field amplitudes in the other regions of the structure, we refer to equations of motions arising from the Hamiltonian in Eq. \eqref{eq:HL}. Working in the Heisenberg picture with Eq. (\ref{eq:HL}), and taking care that the resulting $\psi^{(X)}_{J}(z,t)$ suffer a discontinuity across $z=0$, for each $\psi^{(X)}_{J}(z,t)$ we find that
away from $z=0$ we have \cite{Vernon_lossy_resonators}
\begin{align}\label{eq:prop}
 &\frac{\partial\psi^{(X)}_{J}(z,t)}{\partial t}+v^{(X)}_{J}\frac{\partial\psi^{(X)}_{J}(z,t)}{\partial z}+i\omega_{J}\psi^{(X)}_{J}(z,t)=0,
\end{align}
and at the coupling point we have
\begin{align}
 & \psi^{(X)}_{J>}(0,t)=\psi^{(X)}_{J<}(0,t)-\frac{i\gamma_{J}^{(X)}}{v^{(X)}_{J}}b_{J}(t) \label{eq:channel_eqs}.
\end{align}
We also have
\begin{align}
 & \left(\frac{d}{dt}+\overline{\Gamma}_{J}+i\omega_{J}\right)b_{J}(t)=\sum_X -i\left(\gamma_{J}^{(X)}\right)^{*}\psi^{(X)}_{J<}(0,t),\label{eq:eqwork}
\end{align}
with 
\begin{align}
 & \overline{\Gamma}_{J}=\sum_X{\Gamma_J^{(X)}}, \\
 &\Gamma_J^{(X)} = \frac{\left|\gamma_{J}^{(X)}\right|^{2}}{2v^{(X)}_{J}}. \nonumber
\end{align}
The $\psi^{(X)}_{J<(>)}(0,t)$ in Eqs. \eqref{eq:eqwork} and \eqref{eq:channel_eqs} are introduced to treat the discontinuity at the coupling point;  $\psi^{(X)}_{J<}(0,t)$ is the field $\psi^{(X)}_{J}(z,t)$ for $z<0$ (in-channel), extended
to all $z$ via (\ref{eq:prop}). Likewise, $\psi^{(X)}_{J>}(0,t)$ is the field $\psi^{(X)}_{J}(z,t)$
for $z>0$ (out-channel) extended to all $z$ via (\ref{eq:prop}) \cite{Vernon_lossy_resonators}. From here we impose the conditions in Eqs. \eqref{eq:link} and \eqref{eq:link2} to derive the asymptotic-in and -out field amplitudes.}

\subsection{Asymptotic-in}

{For a general asymptotic-in field associated with in-channel $X$, Eq. \eqref{eq:link} requires that the field in the input region of channel $X$ has the form 
\begin{align}
    \boldsymbol{D}_J(\boldsymbol{r},t)=  \int & dk\: \sqrt{\frac{\hbar \omega_{J}}{2}} \boldsymbol{d}^{(X)}_{J}(x,y) \check{\psi}_{J<}^{(X)}(z,t) e^{i K_J^{(X)}z} \nonumber \\
    & +H.c.,
    \label{eq:asy-in_D}
\end{align}
with 
\begin{align}
    \check{\psi}_{J<}^{(X)}(z,t) = \frac{a_J^{\text{in}(X)}(k)}{\sqrt{2\pi}}e^{i(k-K_J^{(X)})z}e^{-i\omega_{Jk}t}, \label{eq:asyin-Xin}
\end{align}
(c.f. Eqs. \eqref{eq:D_chan_expansion} and \eqref{eq:DChannel}). We take the dependence of $\check{\psi}_{J<}^{(X)}(z,t)$ on $k$ to be understood, and we do not mark it explicitly with a label; the profile $\boldsymbol{d}^{(X)}_{J}(x,y)$ is the waveguide profile $\boldsymbol{d}_{J}(x,y)$ (see Eq. \eqref{eq:DJ}) in waveguide $\bar X$. Eq. \eqref{eq:linka} then requires that we have 
\begin{align}
    \check{\psi}_{J<}^{(Y\neq X)}(z,t) = 0,
\end{align}
so that the field amplitudes are zero in all other input channels. The fields in the output regions of each waveguide have a form similar to Eq. \eqref{eq:asy-in_D}, where the dependence of $\check{\psi}^{(Y)}_{J>}(z,t)$ on $a_J^{\text{in}(X)}(k)$ must be determined. Taking these and seeking a corresponding $\check{b}_{J}(t)=\check{b}_{J}\exp^{-i\omega_{Jk}t}$, we find that the equations (\ref{eq:eqwork}) require 
\begin{align}\label{eq:asy_in_psyY_eom}
 &\left(-i(\omega_{Jk}-\omega_{J})+\overline{\Gamma}_{J}\right)\check{b}_{J}e^{-i\omega_{Jk}t}\\&= \nonumber -i\left(\gamma_{J}^{(X)}\right)^{*} \frac{a_{J}^{\text{in} (X)}(k)}{\sqrt{2\pi}}e^{-i\omega_{Jk}t},\\ 
 &\check{\psi}^{(X)}_{J>}(0,t)=\frac{a_{J}^{\text{in} (X)}(k)}{\sqrt{2\pi}}e^{-i\omega_{Jk}t}-\frac{i\gamma_{J}^{(X)}}{v^{(X)}_{J}}\check{b}_{J}\exp^{-i\omega_{Jk}t},\label{eq:asy_in_psyX_eom}\\
 & \check{\psi}^{(Y\neq X)}_{J>}(0,t)= -\frac{i\gamma_{J}^{(Y\neq X)}}{v^{(Y\neq X)}_{J}}\check{b}_{J}e^{-i\omega_{Jk}t} .
\end{align}
}
Manipulating this we have
\begin{align}
    \left(i(\omega_{J} - \omega_{Jk})+\overline{\Gamma}_{J}\right)\check{b}_{J}&= -i\left(\gamma_{J}^{(X)}\right)^{*} \frac{a_{J}^{\text{in} (X)}(k)}{\sqrt{2\pi}},\\
    \check{b}_J &= \left(\frac{-i\left(\gamma_{J}^{(X)}\right)^{*}}{i(\omega_J - \omega_{Jk}) + \bar{\Gamma_J}}\right)\frac{a_{J}^{\text{{}asy-in} (X)}(k)}{\sqrt{2\pi}}\\
    &= -\sqrt{\frac{\mathcal{L}}{2\pi}} \frac{1}{\sqrt{\mathcal{L}}} \left(\frac{\left(\gamma_{J}^{(X)}\right)^{*}}{(\omega_J - \omega_{Jk}) -i \bar{\Gamma}_J}\right) a_{J}^{\text{in}(X)}(k)\\
    &= -\sqrt{\frac{\mathcal{L}}{2\pi}} \frac{1}{\sqrt{\mathcal{L}}} \left(\frac{\left(\gamma_{J}^{(X)}\right)^{*}}{v_J\left(K^{(X)}_J - k \right) -i \bar{\Gamma}_J}\right) a_{J}^{\text{in} (X)}(k), \label{eq:F_neglect_gvd}\\
    \check{b}_J &= -\sqrt{\frac{\mathcal{L}}{2\pi}} F_{J-}^{(X)}(k) a_{J}^{\text{in} (X)}(k), \label{eq:asyin-b}
\end{align}
with the field enhancement factor $F_{J\pm}^{(X)}(k)$ defined in Eq. \eqref{eq:F}. In Eq. \eqref{eq:F_neglect_gvd}, we have neglected the group velocity dispersion across the resonance; for this to be valid, the group velocity dispersion $\beta_2$ must be small enough that $\frac{1}{v_J}(\omega_J - \omega) >> \beta_2(\omega_J - \omega)^2$. For example, for a 1 GHz resonance linewidth, this would be a good approximation for $\beta_2 < 10^{-20}$ s$^2$/m.

Putting this into Eqs. \eqref{eq:asy_in_psyX_eom} and \eqref{eq:asy_in_psyY_eom}, we have
\begin{align}
 &\check{\psi}^{(X)}_{J>}(0,t)=\frac{a_{J}^{\text{in} (X)}(k)}{\sqrt{2\pi}}e^{-i\omega_{Jk}t}+\frac{i\gamma_{J}^{(X)}}{v^{(X)}_{J}} \sqrt{\frac{\mathcal{L}}{2\pi}} F_{J-}^{(X)}(k) a_{J}^{\text{in} (X)}(k) e^{-i\omega_{Jk}t},\\
 &\check{\psi}^{(X)}_{J>}(0,t)=\left(1 +\frac{i\gamma_{J}^{(X)}}{v^{(X)}_{J}} \sqrt{\mathcal{L}} F_{J-}^{(X)}(k) \right)\frac{a_{J}^{\text{in} (X)}(k)}{\sqrt{2\pi}}e^{-i\omega_{Jk}t}, \label{eq:asyin-Xout} 
\end{align}
and 
\begin{align}
     & \check{\psi}^{(Y\neq X)}_{J>}(0,t)= \frac{i\gamma_{J}^{(Y\neq X)}}{v^{(Y\neq X)}_{J}} \sqrt{\mathcal{L}} F_{J-}^{(X)}(k) \frac{a_{J}^{\text{in}(X)}(k) }{\sqrt{2\pi}} e^{-i\omega_{Jk}t}. \label{eq:asyin-Yout} 
\end{align}
From Eq. \eqref{eq:prop} we have $\check{\psi}^{(Y)}_{J<(>)}(z,t) = \check{\psi}^{(Y)}_{J<(>)}(0,t)e^{i(k-K_J^{(Y)})z}$ for all waveguides $\bar{Y}$, including $\bar{X}$.

We now consider separately each $k$ component of an asymptotic-in field mode $J$. That is, we introduce $\mathcal{D}_{Jk}^{\text{in}(X)}(\boldsymbol{r},t)$ such that 
\begin{align}
    \boldsymbol{D}_J^{\text{in}}(\boldsymbol{r},t) = \sum_{X}\int dk \left(\mathcal{D}_{Jk}^{\text{in}(X)}(\boldsymbol{r},t) + H.c.\right).
\end{align}
The component $\mathcal{D}_{Jk}^{\text{in}(X)}(\boldsymbol{r},t)$ is a piecewise function with the form
\begin{align}
    \mathcal{D}^{\text{in}(X)}_{Jk}(\boldsymbol{r},t)&=\sqrt{\frac{\hbar \omega_{J}}{2}} \boldsymbol{d}^{(Y)}_{J}(x,y) \check{\psi}_{J<(>)}^{(Y)}(z,t) e^{iK_J^{(Y)}z}\\
    &\boldsymbol{r} \in \text{input (output) region of all waveguides $\bar{Y}$},\nonumber\\
    &=\sqrt{\frac{\hbar \omega_{J}}{2}} \frac{\boldsymbol{\mathsf{d}}_{J}(\boldsymbol{r}_{\perp};\zeta)}{\sqrt{\mathcal{L}}}\check{b}_J e^{-i\omega_{Jk} t} e^{i\kappa_{J}\zeta}\\
    &\boldsymbol{r} \in \text{ring}.\nonumber
\end{align}
Using Eqs. \eqref{eq:asyin-Xin}, \eqref{eq:asyin-b}, \eqref{eq:asyin-Xout}, and \eqref{eq:asyin-Yout}, we have 
\begin{align}
    \mathcal{D}^{\text{in}(X)}_{Jk}(\boldsymbol{r},t)&= \sqrt{\frac{\hbar \omega_{J}}{4\pi}} \boldsymbol{d}^{(X)}_{J}(x,y) e^{ikz}{a_{J}^{\text{in} (X)}(k)}e^{-i\omega_{Jk}t} \label{eq:asy-in1}\\ \nonumber &\boldsymbol{r}\in \text{input region of waveguide 
    $\bar{X}$},\\
    &=\sqrt{\frac{\hbar \omega_{J}}{4\pi}} \boldsymbol{d}^{(X)}_{J}(x,y) \left(1 +\frac{i\gamma_{J}^{(X)}}{v^{(X)}_{J}} \sqrt{\mathcal{L}} F_{J-}^{(X)}(k) \right)e^{ikz}{a_{J}^{\text{in} (X)}(k)}e^{-i\omega_{Jk}t} \label{eq:asy-in2} \\ \nonumber &\boldsymbol{r}\in \text{output region of waveguide $\bar{X}$},\\
    &=\sqrt{\frac{\hbar \omega_{J}}{4\pi}} \boldsymbol{d}^{(Y)}_{J}(x,y) \left(\frac{i\gamma_{J}^{(Y\neq X)}}{v^{(Y\neq X)}_{J}} \sqrt{\mathcal{L}} F_{J-}^{(X)}(k) \right) e^{ikz}{a_{J}^{\text{in} (X)}(k)}e^{-i\omega_{Jk}t} \label{eq:asy-in3}\\ \nonumber  &\boldsymbol{r}\in \text{output region of all other waveguides $\bar{Y} \neq \bar{X}$},\\
    &=-\sqrt{\frac{\hbar\omega_{J}}{4\pi}}\boldsymbol{\mathsf{d}}_{J}(\boldsymbol{r}_{\perp},\zeta) F_{J-}^{(X)}(k) e^{i\kappa_{J}\zeta} a_{J}^{\text{in} (X)}(k) e^{-i\omega_{Jk} t} \label{eq:asy-in4}\\ \nonumber  &\boldsymbol{r}\in \text{ring},
\end{align}

Finally, recalling the general form of the asymptotic-in field in Eq. \eqref{eq:D_Asy_expansion}, one has
\begin{align}
 & \mathcal{D}^{\text{in}(X)}_{Jk}(\boldsymbol{r},t)=\boldsymbol{D}_{Jk}^\text{in (X)}(\boldsymbol{r})a_{J}^\text{in (X)}(k)e^{-i\omega_{Jk} t},
\end{align}
and the asymptotic-in field amplitudes listed in Eq. \eqref{eq:in_amplitudes} can be read directly off of Eqs. \eqref{eq:asy-in1}, \eqref{eq:asy-in2}, \eqref{eq:asy-in3}, and \eqref{eq:asy-in4}.

\subsection{Asymptotic-out}

We begin by imposing the appropriate asymptotic behaviour, namely that for an asymptotic-out field associated with an  out-channel $X$, we have
\begin{align}
    \check{\psi}_{J>}^{(X)}(z,t) &= \frac{a_J^\text{out(X)}(k)}{\sqrt{2\pi}}e^{i(k-K_J^{(X)})z}e^{-i\omega_{Jk}t}, \label{eq:asy-out_phi-out}\\
    \check{\psi}_{J>}^{(Y\neq X)}(z,t) &= 0. 
\end{align}
This ensures that the outgoing displacement field in out-channel X has the form of a field propagating through an isolated waveguide, and that this is the system's only outgoing field. These boundary conditions along with Eqs. \eqref{eq:channel_eqs} and \eqref{eq:eqwork} give
\begin{align}
    &\frac{a_J^\text{out(X)}(k)}{\sqrt{2\pi}}e^{-i\omega_{Jk}t} = \check{\psi}_{J<}^{(X)}(0,t) - \frac{i\gamma_{J}^{(X)}}{v^{(X)}_{J}}\check{b}_{J}(t) \label{eq:asy_out_X_eom}\\
    &0 = \check{\psi}_{J<}^{(Y\neq X)}(0,t) - \frac{i\gamma_{J}^{(Y\neq X)}}{v^{(Y\neq X)}_{J}}\check{b}_{J}(t) \label{eq:asy_out_Y_eom}\\
    &\left(\frac{d}{dt}+\overline{\Gamma}_{J}+i\omega_{J}\right)\check{b}_{J}(t)=-i\left(\gamma_{J}^{(X)}\right)^{*}\psi^{(X)}_{J<}(0,t) - i\sum_{Y\neq X}\left(\gamma_{J}^{(Y\neq X)}\right)^{*}\psi^{(Y\neq X)}_{J<}(0,t),  \label{eq:asy_out_b_eom}
\end{align}
 
Putting $\check{b}_J(t) = \check{b}_J e^{-i\omega_{Jk}t}$ and rearranging Eqs. \eqref{eq:asy_out_X_eom} and \eqref{eq:asy_out_Y_eom}, we have 
\begin{align}
    & \check{\psi}_{J<}^{(X)}(0,t) = \frac{a_J^\text{out(X)}(k)}{\sqrt{2\pi}}e^{-i\omega_{Jk}t} + \frac{i\gamma_{J}^{(X)}}{v^{(X)}_{J}}\check{b}_{J}e^{-i\omega_{Jk}t} \label{eq:asy_out_X_eom2} \\
    &\check{\psi}_{J<}^{(Y\neq X)}(0,t) =  \frac{i\gamma_{J}^{(Y\neq X)}}{v^{(Y\neq X)}_{J}}\check{b}_{J}e^{-i\omega_{Jk}t}.\label{eq:asy_out_Y_eom2}
\end{align}
Putting these into Eq. \eqref{eq:asy_out_b_eom} and using $\Gamma_J^{(Y)} = \frac{|\gamma_J^{(Y)}|^2}{2v_J^{(Y)}}$, we have 
\begin{align}
 & \left(-i(\omega_{Jk} - \omega_{J})+\overline{\Gamma}_{J}\right)\check{b}_{J}e^{-i\omega_{Jk}t}= -i\left(\gamma_{J}^{(X)}\right)^{*} \frac{a_{J}^{\text{out} (X)}(k)}{\sqrt{2\pi}} e^{-i\omega_{Jk}t} + 2\Gamma_J^{(X)} \check{b}_J e^{-i\omega_{Jk}t} + \sum_{Y \neq X} 2\Gamma_J^{(Y)} \check{b}_J e^{-i\omega_{Jk}t},
\end{align}
and since $\overline{\Gamma}_J = \sum_{X} \Gamma_J^{(X)}$, this rearranges to 
\begin{align}
 & \left(-i(\omega_{Jk} - \omega_{J})-\overline{\Gamma}_{J}\right)\check{b}_{J}= -i\left(\gamma_{J}^{(X)}\right)^{*} \frac{a_{J}^{\text{out} (X)}(k)}{\sqrt{2\pi}},
\end{align}
so 
\begin{align}
    \check{b}_J &= \left(\frac{-i\left(\gamma_{J}^{(X)}\right)^{*}}{i(\omega_J - \omega_{Jk}) - \bar{\Gamma_J}}\right)\frac{a_{J}^{\text{out} (X)}(k)}{\sqrt{2\pi}}\\
    &= -\sqrt{\frac{\mathcal{L}}{2\pi}} \frac{1}{\sqrt{\mathcal{L}}} \left(\frac{\left(\gamma_{J}^{(X)}\right)^{*}}{v_J\left(K^{(X)}_J - k \right) +i \bar{\Gamma}_J}\right) a_{J}^{\text{out} (X)}(k),\\
    \check{b}_J &= -\sqrt{\frac{\mathcal{L}}{2\pi}} F_{J+}^{(X)}(k) a_{J}^{\text{out}(X)}(k).\label{eq:asy-out_b}
\end{align}

Using this in Eqs. \eqref{eq:asy_out_X_eom2} and \eqref{eq:asy_out_Y_eom2}, we have 
\begin{align}
 &\check{\psi}^{(X)}_{J<}(0,t) = \left( 1  - \frac{i\gamma_{J}^{(X)}}{v^{(X)}_{J}} \sqrt{\mathcal{L}} F_{J+}^{(X)}(k)  \right) \frac{a_{J}^{\text{out} (X)}(k)}{\sqrt{2\pi}}e^{-i\omega_{Jk}t}\label{asy-out_phiX}\\
 & \check{\psi}^{(Y\neq X)}_{J<}(0,t)= -\frac{i\gamma_{J}^{(Y\neq X)}}{v^{(Y\neq X)}_{J}}\sqrt{\mathcal{L}} F_{J+}^{(X)}(k) \frac{a_{J}^{\text{out}(X)}(k)}{\sqrt{2\pi}} e^{-i\omega_{Jk}t}  \label{asy-out_phiY}.
\end{align}
From here the derivation of the field amplitudes follows the asymptotic-in case; one introduces the components $\mathcal{D}^{\text{out}(X)}_{Jk}(\boldsymbol{r},t)$ of the full asymptotic-out field, and finds that the amplitudes given in Eq. \eqref{eq:out-amplitude} follow from Eqs. \eqref{asy-out_phiX}, \eqref{asy-out_phiY}, \eqref{eq:asy-out_b}, and \eqref{eq:asy-out_phi-out}.

\newpage

\section{Add-drop sample calculation: Pair rates with varying coupling to waveguides.}
\label{section:supplementary}

\begin{figure}[h!]
    \centering
    \includegraphics[scale=1.3]{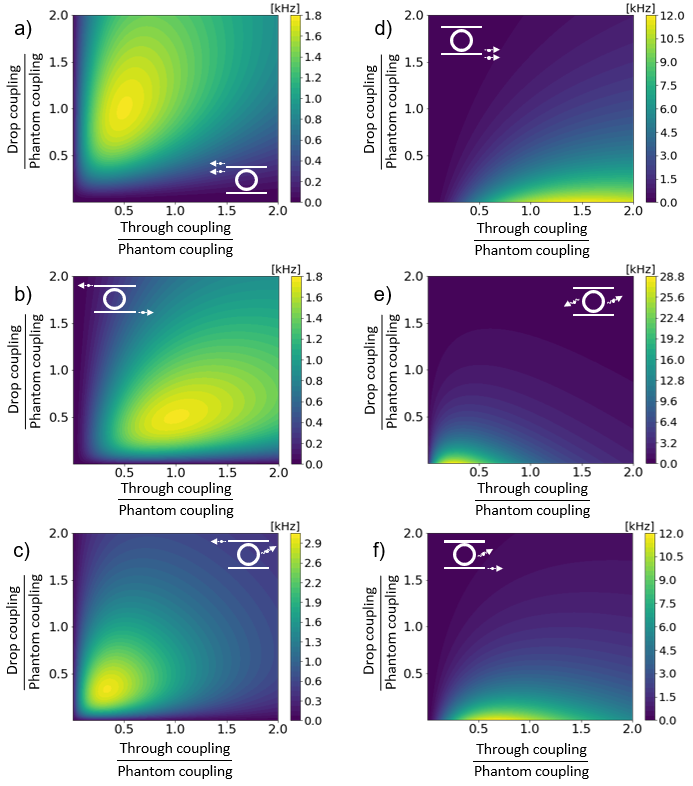}
    \caption{Rate of signal and idler photons in channels $X$ and $X'$ respectively for the add-drop system, with variable coupling to the `Through' and `Drop' waveguides ($\Gamma^{(T)}$ and $\Gamma^{(D)}$) for a) $X=D,\:X'=D$; b) $X=D,\:X'=T$ or $X=T,\:X'=D$; c)  $X=D,\:X'=P$ or $X=P,\:X'=D$; d) $X=T,\:X'=T$; e) $X=P,\:X'=P$; f) $X=P,\:X'=T$ or $X=T,\:X'=P$.}
    \label{fig:add-drop_appendix}
\end{figure}

In Sec. \ref{section:add-drop} we discussed photon pair generation in an add-drop structure. In particular, we discussed the rate of photons in different sets of channels, including the phantom channel, and their dependence on the ring's coupling to the through and drop ports. This dependence is plotted in Fig. \ref{fig:add-drop_appendix} for all the different trajectories for the photon pairs.

\end{document}